\documentclass[eng]{ajceam-class}

\usepackage{chemformula}
\usepackage{multirow}
\usepackage{algpseudocode}
\usepackage{float}
\usepackage{amsmath}
\usepackage{makecell}
\usepackage{listings}
\usepackage{stfloats}
\usepackage{hyperref}
\usepackage{afterpage}

\usepackage{float}
\usepackage{physics}

\title{Extracting Structured Seed-Mediated Gold Nanorod Growth Procedures from Literature with GPT-3}

\author[1]{Nicholas Walker}
\author[1,4]{John Dagdelen}
\author[2,4]{Kevin Cruse}
\author[1,4]{Sanghoon Lee}
\author[1,5]{Samuel Gleason}
\author[1,4]{Alexander Dunn}
\author[2,4]{Gerbrand Ceder}
\author[2,4,5,6]{A. Paul Alivisatos}
\author[3,4,6]{Kristin A. Persson}
\author[1]{Anubhav Jain}

\affil[1]{Energy Technologies Area, Lawrence Berkeley National Laboratory, 1 Cyclotron Road, Berkeley, 94720, CA, United States of America}
\affil[2]{Materials Sciences Division, Lawrence Berkeley National Laboratory, 1 Cyclotron Road, Berkeley, 94720, CA, United States of America}
\affil[3]{Molecular Foundry, Lawrence Berkeley National Laboratory, 1 Cyclotron Road, Berkeley, 94720, CA, United States of America}
\affil[4]{Department of Materials Science and Engineering, University of California Berkeley, 210 Hearst Memorial Mining Building, Berkeley, 94720, CA, United States of America}
\affil[5]{Department of Chemistry, University of California Berkeley, 419 Latimer Hall, Berkeley, 94720, CA, United States of America}
\affil[6]{Kavli Energy NanoScience Institute, University of California Berkeley, 101C Campbell Hall, Berkeley, 94720, CA, United States of America}

\firstauthor{Walker}

\contactauthor{Anubhav Jain}
\email{ajain@lbl.gov}

\thisvolume{XX}
\thisnumber{XX}
\thismonth{Month}
\thisyear{20XX}

\abstract{Although gold nanorods have been the subject of much research, the pathways for controlling their shape and thereby their optical properties remain largely heuristically understood. Although it is apparent that the simultaneous presence of and interaction between various reagents during synthesis control these properties, computational and experimental approaches for exploring the synthesis space can be either intractable or too time-consuming in practice. This motivates an alternative approach leveraging the wealth of synthesis information already embedded in the body of scientific literature by developing tools to extract relevant structured data in an automated, high-throughput manner. To that end, we present an approach using the powerful GPT-3 language model to extract structured multi-step seed-mediated growth procedures and outcomes for gold nanorods from unstructured scientific text. GPT-3 prompt completions are fine-tuned to predict synthesis templates in the form of JSON documents from unstructured text input with an overall accuracy of $86\%$. The performance is notable, considering the model is performing simultaneous entity recognition and relation extraction. We present a dataset of 11,644 entities extracted from 1,137 papers, resulting in 268 papers with at least one complete seed-mediated gold nanorod growth procedure and outcome for a total of 332 complete procedures.}
 

\begin{document}

\maketitle
\printcontactdata

\section{Introduction}\label{introduction}

Gold nanoparticles have been synthesized for centuries due to their interesting optical properties, dating back to the Lycurgus Cup from 4\textsuperscript{th} century Rome,\cite{aunp_hist_0} as well as imperial bowls and decorated dishes from the Qing dynasty.\cite{aunp_hist_1} However, scientific interest did not develop until the work of Michael Faraday in the mid-19\textsuperscript{th} century, when he accidentally synthesized colloidal gold while investigating the interaction between light and matter.\cite{aunp_hist_2} In the last three decades, chemists have developed the ability to synthesize anisotropic metal nanoparticles in a controllable and reproducible fashion.\cite{aunr_hist} Around the turn of the millennium, multi-step seed-mediated growth methods were developed to prepare gold nanorods with aspect ratios ranging from $8$ to $20$.\cite{aunr_hist,aunr_synth,aunr_synth_2} This generated a great deal of interest in anisotropic gold nanoparticles due to a combination of the convenience of the wet-chemistry approach, as well as the ability to tune the shape of the synthesized nanorods. The anisotropic gold nanoparticles, in turn, provide access to shape-dependent optical phenomena not observed with spherical gold nanoparticles.\cite{aunr_optical_0,aunr_optical_1,aunr_optical_2,aunr_optical_3} Their applications are widespread across many domains, including semiconductor technology,\cite{aunp_semiconductor,aunp_semiconductor_2} biomedicine,\cite{aunp_biomedical,aunp_biomedical_2} and cosmetics.\cite{aunp_cosmetics} The suitability of a nanoparticle for a particular application depends on its morphology and size, which correspond to different plasmonic properties.\cite{aunp_size_morph,aunp_size,aunp_size_shape}

Despite the popularity of anisotropic gold nanoparticles, systematic investigation of the control of these properties has only recently been approached.\cite{aunp_syst} Although some theories and models do exist for identifying and explaining the mechanisms of synthesis that determine nanoparticle morphology,\cite{aunr_hist,aunp_form_model,aunp_shape_control_0,aunp_shape_control_1} most synthesis exploration is still guided by heuristics based on domain knowledge.

For gold nanorods, it is clear that the simultaneous presence of various reagents during the synthesis affects the characteristics of the resulting gold nanoparticles.\cite{aunr_hist} To better understand these effects, computational simulation and analysis of the formation energetics of the nanoparticles or the nucleation and growth steps can be used. Density functional theory (DFT) can be used to investigate the energetic landscape of potential gold nanoparticle morphologies, including the effects of surface ligands that are vital for the solution-phase synthesis of noble metal nanoparticles.\cite{aunp_dft,aunp_ligand_0,aunp_ligand_1} However, this approach does not account for the nuances of nucleation and growth competition in solution-based nanoparticle syntheses. These aspects can be addressed by modeling real-time growth and dispersity dynamics with continuum-level model, though this sacrifices access to small-scale energetics granted by DFT.\cite{aunp_continuum} Alternatively, direct experimentation can be used to explore the synthesis space by varying precursor amounts over many experiments, though this is impractical due to the both the number of experiments required to sample the synthesis space and the condition that a single experiment can take many hours to complete. Automated labs may address this problem in the future, though most are still in their infancy.

A third approach seeks to leverage the wealth of information contained in scientific literature. Many seed-mediated gold nanorod recipes have been published in the materials science and chemistry literature, but parsing them requires domain experts to manually read these articles to retrieve the relevant precursors, procedures, laboratory conditions, and target characterizations. This comes with its own complications, however, as over time, the body of materials science literature has grown to an unwieldy extent, preventing researchers from absorbing the full breadth of information contained in established literature or even reasonably following research progress as it emerges.\cite{nlp_matsci_mining_1} Thus, it is unreasonable to expect domain experts in gold nanoparticle synthesis to manually read and parse the complete existing synthesis literature efficiently, motivating the development of high-throughput text-mining methods to extract this information.

The resulting databases built with these methods are the first steps toward developing data-driven approaches to understanding synthesis, which are being developed at an accelerating pace as a rapidly emerging third paradigm of scientific investigation. Generally speaking, these approaches involve the use of both conventional and machine learning methods to both build large databases and perform downstream analysis and inference over said databases. Natural language processing (NLP) has been successfully applied in the chemical, medical, and materials sciences to produce structured data from unstructured text using methods and models such as pattern recognition, recurrent neural networks, and language models.\cite{matsynth_corpus,ner_chemistry_2,ner_chemistry_3,ner_medical,ner_biomedical,bioelectra,weston_ner,matsynth_corpus, nlp_matsci_mining_0,nlp_matsci_0,nlp_matsci_1,nlp_matsci_2,nlp_matsci_3,nlp_matsci_4,nlp_chem_transfer,nlp_chem_ner_0,nlp_chem_ner_1,nlp_chem_ner_2,nlp_chem_ner_3,nlp_chem_ner_4,nlp_chem_ner_5,nlp_biomed_ner_0,matbert_ner}

For applications specifically related to materials synthesis, data-driven approaches have been successful for tasks such as materials discovery, synthesis protocol querying, and simulation and interpretation of characterization results.\cite{tm_mat_disc_0,tm_mat_disc_1,tm_protocol,tm_char_0,tm_char_1} However, these approaches are fundamentally limited by the quality of the data, such as the completeness and substance of the data source. To address this, careful data curation is necessary, as seen with the construction and maintenance of large databases of characteristic features of nanostructures.\cite{nano_db}

Recently, the wealth of unstructured information about gold nanoparticle synthesis and characterization in literature has been directly tapped through the combination of various NLP models and other text-mining techniques to produce a dataset of over five thousand codified gold nanoparticle synthesis protocols and outcomes.\cite{aunp_synth_dataset} This general dataset contains a wealth of information, including detected materials, material quantities, morphologies, synthesis actions, and synthesis conditions, as well as tags for seed-mediated synthesis, synthesis paragraph classifications, and characterization paragraph classifications.

Despite the breadth of accurate information provided, the general dataset still suffers from a few pitfalls: (i) the inability to distinguish between seed and growth solution procedures in seed-mediated growth synthesis; (ii) the inability to detect references to materials that do not contain specific formulae or chemical names (e.g. ``AuNP seed solution''); and (iii) the inability to detect target morphologies as opposed to incidentally mentioned morphologies. To address these issues, this work intends to use a large sequence-to-sequence language model to extract full synthesis procedures and outcomes in a single-step inference. Generally speaking, a sequence-to-sequence model in NLP maps an input sequence to an output sequence by learning to produce the most likely completion of the input by conditioning the output on the input.\cite{s2s}

In this work, we leverage the capabilities of the latest language model in the Generative Pre-trained Transformer (GPT) family, GPT-3,\cite{gpt3} to build a dataset of highly structured synthesis templates for seed-mediated gold nanorod growth. A similar approach using GPT-3 to build materials science datasets has been applied to extracting dopant-host material pairs, cataloging metal-organic frameworks, and extracting general chemistry/phase/morphology/application information for materials.\cite{gpt3_mastci} We extracted these templates for seed-mediated gold nanorod growth from 2,969 paragraphs across 1,137 papers, starting with using a question-answering framework aided by the zero-shot performance of GPT-3 to construct a small initial dataset. We then fine-tuned GPT-3 to produce complete synthesis templates for input paragraphs. Fine-tuning GPT-3 consists of using multiple examples of paragraph and synthesis template pairs to train GPT-3 to perform this specific task. Each synthesis template in the final dataset contains information on relevant synthesis precursors, precursor amounts, synthesis conditions, and characterization results, all structured in a JSON format. This dataset provides reproducible summaries of procedures and outcomes, explicitly establishing the relationships between the components of the recipe (e.g. accurately linking the correct volumes and concentrations with the correct precursors in the correct solution). However, this specificity comes at the cost of generality, as the dataset focuses on seed-mediated gold nanorod growth. The final dataset consists of 11,644 entities extracted from 1,137 papers, 268 of which contain least one complete seed-mediated gold nanorod growth procedure and outcome for a total of 332 complete procedures.

\section{Dataset}\label{dataset}

The relevant data for constructing the training, testing, and prediction data for this model was collected using the database of gold nanoparticle synthesis protocols and outcomes developed by Cruse et al.\cite{aunp_synth_dataset} from the full-text database developed by Kononova et al.\cite{matsynth_corpus} through text- and data-mining agreements with several major scientific journal publishers. The original full-text database contains more than 4.9 million materials science articles, and the pipeline for identifying and extracting gold nanoparticle synthesis articles consists of progressively finer-meshed filtering steps using text-mining tools adapted from Kononova et al.\cite{matsynth_corpus} and Wang et al.\cite{sol_synth_dataset} These steps include regular expression matching to identify nanomaterial papers, document and vocabulary vectorization using term frequency-inverse document frequency (TF-IDF) to reveal papers related more to gold than other noble metals, BERT-based binary classifiers to identify paragraphs related to gold nanoparticle synthesis or characterization (particularly morphological information), a combination of BiLSTM-based named entity recognition (NER) and rules-based methods to extract synthesis procedure details from synthesis paragraphs, and MatBERT\cite{matbert_ner} NER to extract morphology and size information from characterization paragraphs.

Using the extracted information, 5,145 papers were identified to contain gold nanoparticle synthesis protocols,\cite{aunpdataset} of which 1,137 papers were found to contain seed-mediated recipes using the "seed\_mediated" flag as well as rod-like morphologies ("rod or "NR" in "morphologies" under "morphological\_information") or aspect ratio measurements ("aspect" or "AR" in "measurments" under "morphological\_information"). This was done to filter the total papers down to only those likely to contain seed-mediated synthesis recipes for gold nanorods.

\section{Methods}\label{methods}

At the core of the GPT-1 model was a focus on improving language understanding by generative pre-training involving the use of a large language model in conjunction with a very large and diverse pre-training corpus with long stretches of contiguous text, which greatly facilitated the model's ability to learn ``world knowledge'' alongside its ability to process long-range dependencies.\cite{gpt1} For a sequence-to-sequence generative model, outputs are generated by maximizing the log probability of $p(\textrm{output}|\textrm{input})$.\cite{s2s} To further improve zero-shot performance for both learning and task transfer, GPT-2 modified the training objective to include task conditioning, $p(\textrm{output}|\textrm{input},\textrm{task})$, thus establishing the model as an unsupervised multitask learner.\cite{gpt2} With GPT-3, more extensions of the model size and the pre-training corpus have produced a model with considerable capacity for few-shot learning that is capable of producing text that is difficult to distinguish from human-written text or performing tasks it was not explicitly trained on, such as writing code or summing numbers.\cite{gpt3} We employed the 175 billion parameter variant of GPT-3 (OpenAI Davinci) for this work.

Of the 1,137 papers identified to contain information about seed-mediated gold nanorod synthesis, 240 (consisting of 661 relevant paragraphs) were randomly sampled and fully annotated with JSON-formatted recipes by a single annotator with machine assistance to serve as a training set. An additional 40 papers (consisting of 117 paragraphs) were used for prediction and corrected to serve as a testing set. Each relevant paragraph was separately annotated due to length constraints imposed by GPT-3, which limits the capability to process an entire article at once. A limit of 2,048 tokens is shared between the input prompt and the output completion, corresponding to approximately 1,500 words.\cite{gpt3}

\subsection{Overall Procedure}

\begin{figure}[H]
\centering
\includegraphics[width=\linewidth]{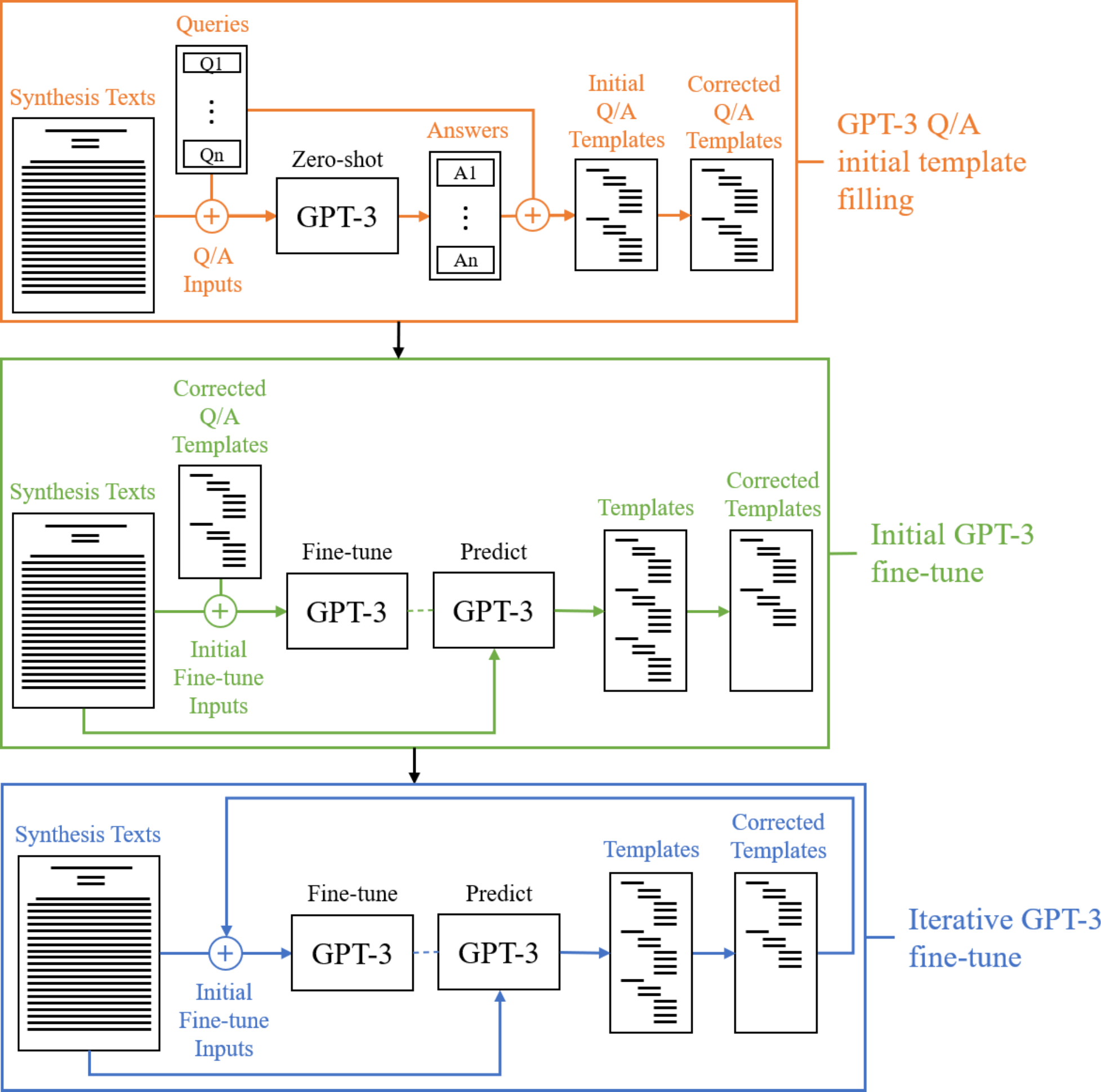}
\caption{A diagram illustrating the overall procedural approach for extracting synthesis templates from text with GPT-3 is shown. All unstructured text paragraphs were drawn from the seed-mediated gold nanorod growth dataset of 1,137 papers (purple). The first stage involves filling initial templates using a zero-shot question/answer framework with GPT-3, which is then corrected (orange). The plus sign indicates a combination of the texts and queries used as input. Template correction is done through manual editing of the templates. These corrected templates are used to fine-tune an initial GPT-3 model, which produces complete templates in a single prediction (green). From there, the process of iteratively predicting more templates with a fine-tuned model, correcting them, adding them to the training set, and then fine-tuning the model again is then performed (blue). The plus signs for these stages indicate that text-template pairs are used as input for fine-tuning.}
\label{fig:procedure}
\end{figure}

A diagram outlining the general process for producing the final fine-tuned model for template-filling is shown in Figure \ref{fig:procedure}. In the initial stage (orange), a simple question-answering framework is used to individually fill in templates for an initial set of paragraphs. These results are then corrected and used as an initial training set for fine-tuning GPT-3 to produce complete templates in the second stage (green). The final stage is an iterative training process in which new templates are predicted, corrected, and added to the training set to update the fine-tuned model, thus improving its performance with each iteration. All input texts for each stage are drawn from the original dataset of synthesis text filtered down to paragraphs likely to describe seed-mediated gold nanorod growth (purple). Default settings through the OpenAI API are used for all fine-tunes of the GPT-3 Davinci model, and a temperature of zero is used for all model predictions with a double line break as the stop sequence.

\subsection{Template Structure and Annotation Scheme}

\begin{figure}[H]
\centering
\includegraphics[width=.9\linewidth]{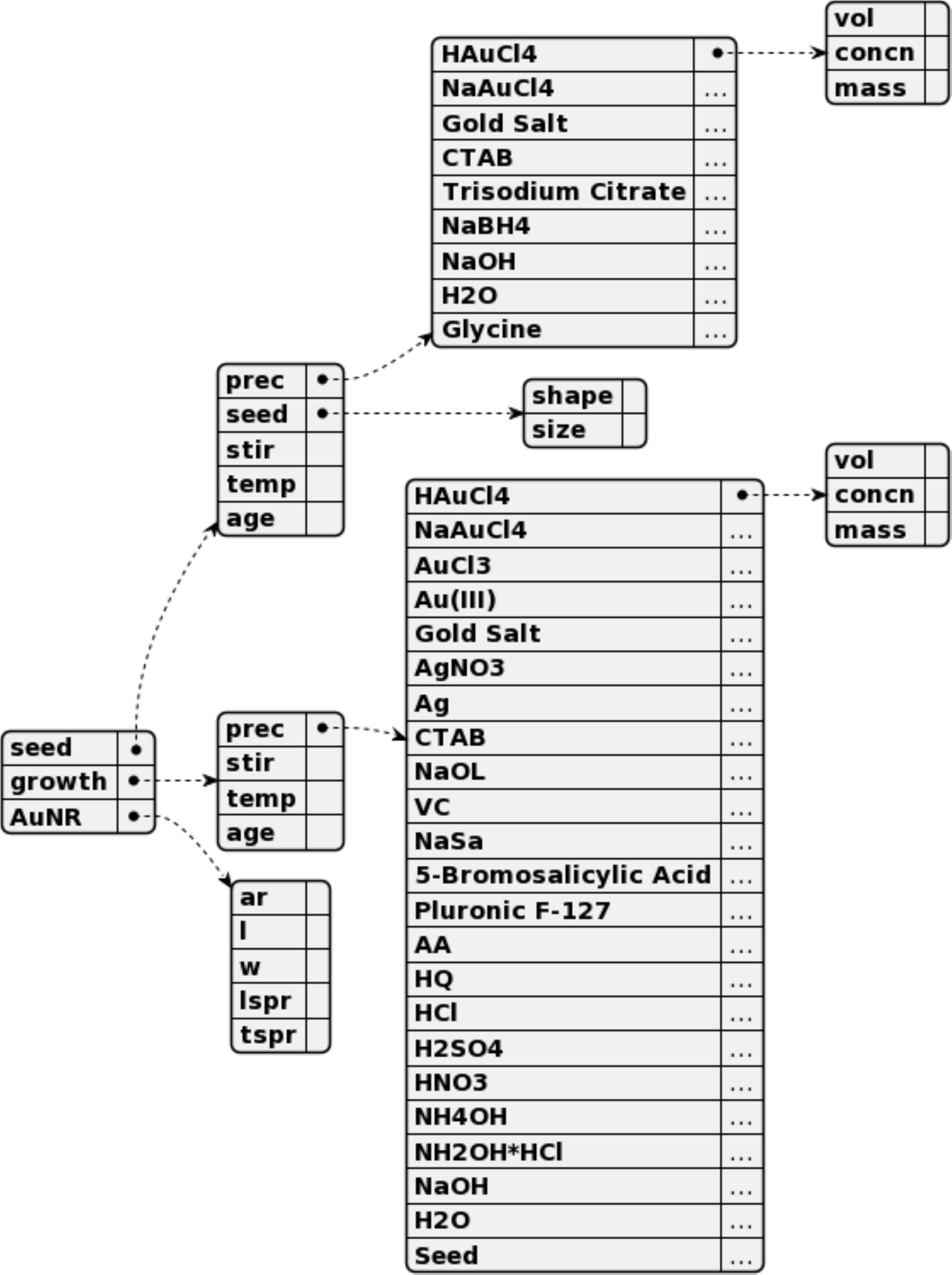}
\caption{A diagram representing the structure of the seed-mediated gold nanorod growth JSON template. From left to right, the structure is divided into three components, the seed solution, the growth solution, and the resulting gold nanorods. For the seed and growth solution components, there are entries for the precursors and their associated quantities, as well as entries for experimental conditions such as the age and aging temperatures of the solutions and stir rates when adding the reducing agent (for the seed solution) or the seed solution (for the growth solution). For the gold nanorod component, there are entries for the characterization information that may be present, including the aspect ratio (ar), length (l), and longitudinal/transverse surface plasmon resonances (l/tspr).}
\label{fig:template}
\end{figure}

The structure and content of the synthesis templates are shown in Figure \ref{fig:template}. The synthesis templates are stored as JSON documents, which contain three components: the seed solution, the growth solution, and the resulting nanorods. For the seed and growth solutions, the precursors and their associated volumes (vol), concentrations (concn), and/or masses are recorded, as well as the ages of the respective mixed solutions at the time of use and the temperatures (temp) at which they are aged. Furthermore, the stirring rates when adding sodium borohydride (NaBH4) to the seed solution and when adding the seed solution to the growth solution are recorded. The shape and size of the gold seeds in the seed solution are also noted. For the gold nanorods (AuNR), the aspect ratios (ar), lengths (l), widths (w), and longitudinal/transverse surface plasmon resonances (SPRs) are recorded. The JSON documents have identical structures and thus contain an entry for every value that can be requested; any values not present in a given paragraph are filled with an empty string.

When available, numerical quantities with units are extracted. For precursor volumes, the units are provided in variations of liters, though the concentrations may be measured in either molarity, molality, or weight percentage. In some cases, the total volume of a collection of precursors may be specified instead of the individual volumes of the precursors. In this case, the explicit volume is associated with the first precursor and the volumes for the remaining precursors refer to the name of the first precursor, implicitly communicating a shared volume. For temperatures, degrees Celsius are most commonly provided, though more qualitative descriptions such as ``room temperature'' will still be recorded if the explicit temperature is not provided in the text but a qualitative description is. Similarly, for solution ages, minutes or hours are most common, but sometimes only descriptions like ``overnight'' are provided and recorded. For stirring rates, the revolutions per minute (rpm) is preferred, but many papers will instead provide descriptions such as ``gentle'' or ``vigorous'' that are recorded. For the gold nanorod properties, aspect ratios are unitless while the other quantities (length, width, SPRs) are provided in units of length, with the exception of some cases where the LSPR is only provided as ``NIR'' (near-infrared). Throughout all stages of the annotation process, three additional researchers were consulted to reach a consensus on the appropriate annotations for various edge cases caused by unclear wording or other ambiguities.

\subsection{Question Answering Completions}

\begin{figure}[H]
\centering
\includegraphics[width=\linewidth]{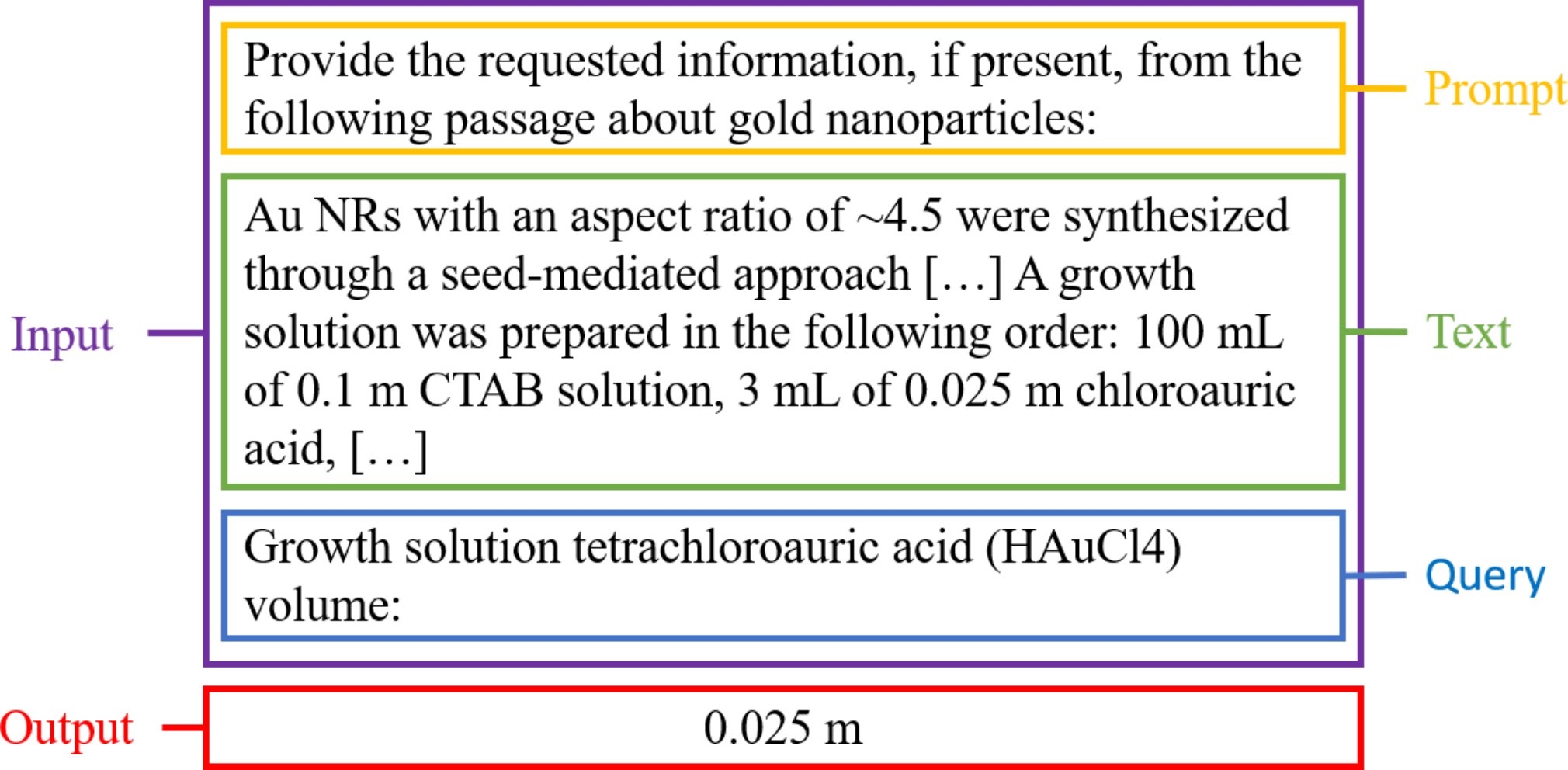}
\caption{An example of a question answering completion using GPT-3. The input is bounded by a purple box containing the prompt (orange), paragraph text (green), and query (blue). The output is bounded by a red box.}
\label{fig:qa_example}
\end{figure}

Unfortunately, the standard pre-trained GPT-3 Davinci model is not capable of providing consistent completed templates of high quality in one request. However, the model is capable of answering simple questions about synthesis paragraphs without any fine-tuning, which allows for the fields of the synthesis templates to be individually filled using answers from a simple question-answering framework using GPT-3. An example is shown in Figure \ref{fig:qa_example}.\cite{example_text} This machine-assisted annotation approach avoids the laborious process of manually filling in each field of the templates by hand, as an annotator only needs to verify and correct the provided answers as-needed. However, this approach does not scale well to large numbers of papers, as each query is a separate model request, meaning that each paragraph in each paper would require a large number of requests in order to fill a single template. Therefore, this approach is used to construct an initial dataset consisting of synthesis templates for paragraphs from a small number of papers. Due to the small number of papers used, this initial dataset does not necessarily capture the variety of precursors or manners in which critical data can be communicated in text. Nevertheless, these initial templates, when corrected, provide a suitable starting point for fine-tuning GPT-3 to provide complete synthesis templates in single requests for each paragraph. Through an iterative process of fine-tuning GPT-3 on the available templates, predicting new templates, correcting them, and fine-tuning a new model using all of the corrected templates constructed thus far, a final fine-tuned model can be obtained.

The initial synthesis template dataset was constructed using the zero-shot question-answering framework with 40 randomly sampled papers. If a relevant precursor, condition, or characterization was identified with regular expression pattern matching in the paragraph, the framework would be to request the information using GPT-3. For example, if ``ascorbic acid'', ``AA'', ``vitamin C'', or ``C6H8O6'' appeared in the paragraph, the framework would request the volume, concentration, and mass of ascorbic acid. This initial dataset only requested information about the eight most common precursors, including ``HAuCl4'', ``CTAB'', and ``NaBH4'' for the seed solution, and ``HAuCl4'', ``CTAB'', ``AgNO3'', ``AA'', and ``seed solution'' for the growth solution. To capture different ways of expressing each precursor, multiple aliases were checked to include variations on chemical names as well as the chemical formulas. Additionally, the framework requested information about the stir rate when adding NaBH\textsubscript{4} to the seed solution, the age of the seed solution, the temperature of the seed solution during aging, the size and shape of the seeds, the stir rate when adding the seed solution to the growth solution, the age of the growth solution, and the temperature of the growth solution during aging. All request completions for each paragraph were aggregated into a single JSON entry according to the synthesis template scheme shown in Figure \ref{fig:template}.

The approach of using zero-shot GPT-3 question answering requests to fill the templates tended to produce poor results, but it offered an acceptable starting point for collecting structured recipes. Most of the templates only required correcting the incorrect entries, rather than filling them in manually from scratch, which greatly accelerated the creation of the initial dataset. However, some entries had to be added from scratch due to recipes including precursors outside the initial set of eight common precursors. Note that the static nature of the synthesis templates across all paragraphs means that when one paragraph requires the addition of a new precursor to the template, this is applied to all templates for all paragraphs. Additionally, annotation was done strictly, requiring that the synthesis method must be seed-mediated growth and the target gold nanoparticle morphology must be nanorods. This provides an important test for the model, as the difference between recipes that produce very similar morphologies can sometimes be subtle.

\subsection{Fine-tuning Procedure and Dataset Construction}

These corrected templates derived from the question answering completions provided an initial training set for fine-tuning GPT-3 to produce the desired filled templates. From there, templates for paragraphs from 40 more randomly sampled papers were iteratively predicted, corrected (adding new precursors as necessary), and added to the training set until templates for paragraphs from 240 papers had been corrected in total. With each iteration, the correction process became much easier and faster. Initially, templates for information-dense paragraphs took approximately 4 minutes to validate and correct, whereas, by the end of the process, they took around a minute each. This is because GPT-3 largely predicted filled templates with high accuracy. The testing dataset was composed of paragraphs from an additional random sampling of 40 papers. Not all of the papers filtered from the original dataset were guaranteed to contain information that should be placed into synthesis templates. For example, seed-mediated growth or nanorod measurements and morphologies may only be incidentally mentioned in a given paragraph that is otherwise not relevant to a specific seed-mediated gold nanorod growth procedure. Of the 240 papers in the training set and the 40 papers in the testing set, 141 and 23 papers respectively contained at least one paragraph with information that could be placed into a synthesis template.

\section{Results}\label{results}

The described training dataset of synthesis templates was used to fine-tune a GPT-3 model to reproduce said synthesis templates from the unstructured text. Default parameters for the fine-tuning process were employed, incurring a cost of 85.30 USD. The predictions over the testing dataset (40 papers composed of 117 paragraphs) took around eighty minutes to complete and incurred a cost of 14.39 USD. The performance of the fine-tuned model was then evaluated using the corrected testing dataset.

\subsection{Error Evaluation Examples and Definitions}

\begin{figure}[H]
\centering
\includegraphics[width=\linewidth]{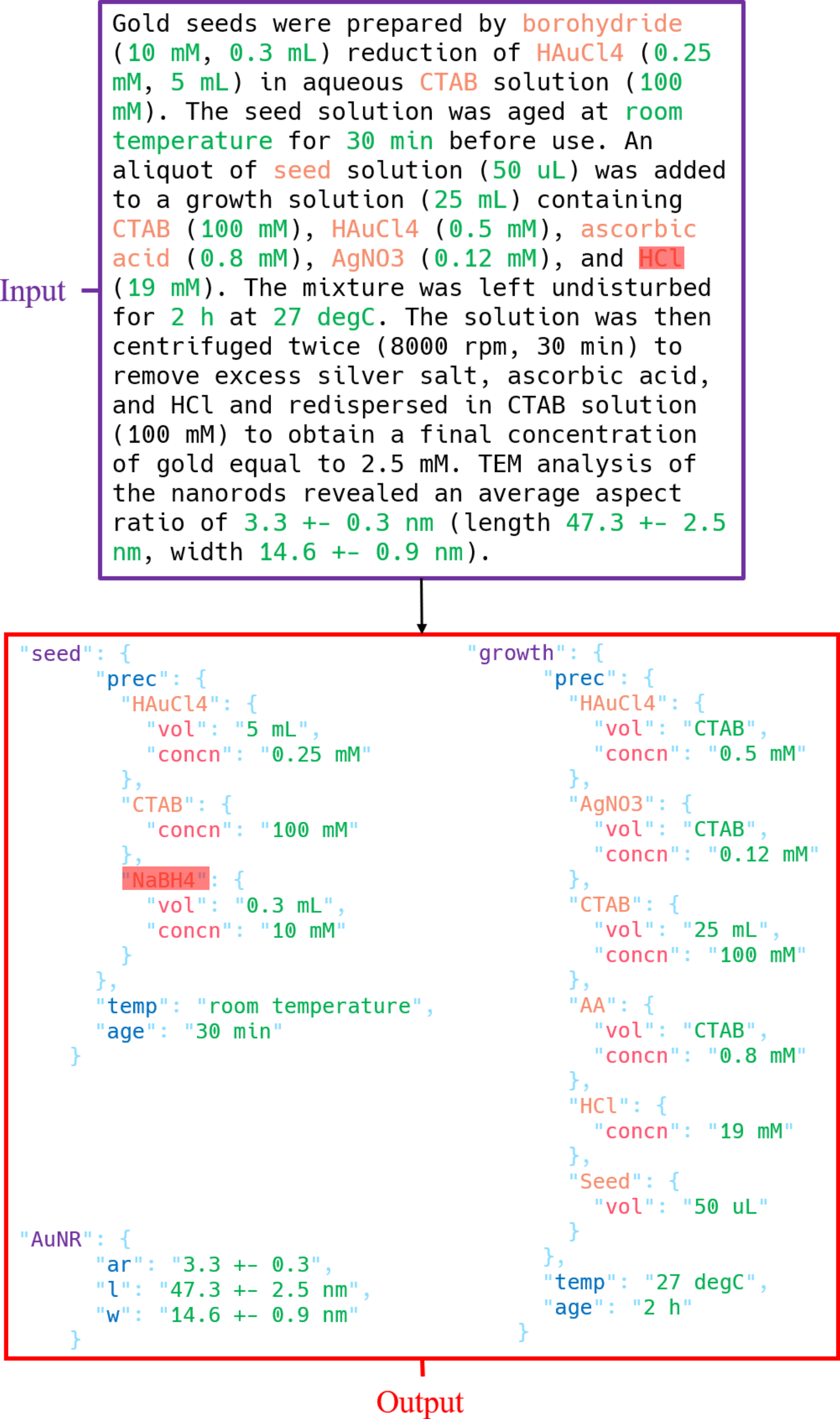}
\caption{A model prediction example is shown, with empty entries omitted. The original unstructured text is shown on the left, and the components of the predicted synthesis template in JSON form are shown on the right. The important information from the unstructured text is colored in orange (for precursors) and green (for quantities), while any errors are highlighted in red.}
\label{fig:static_template_example_prediction}
\end{figure}

An example prediction is depicted in Figure \ref{fig:static_template_example_prediction}.\cite{example} Errors are highlighted in red. For this example, two errors were made. First, the quantities for ``Borohydride'' in the seed solution were instead placed under ``NaBH4'' in the seed solution. Arguably, this is not truly an error since sodium borohydride is often conventionally referred to as ``borohydride'', possibly indicating ``world knowledge'' exhibited by GPT-3. However, there are technically other borohydrides, such as potassium borohydride, that can be used as a reducing agent for seed-mediated gold nanorod growth,\cite{synth_kbh4} so this was still marked as incorrect due to possible ambiguity. The second error was the failure to extract the HCl volume. Note the rather complex relationship in the growth solution precursor volumes, where CTAB, HAuCl\textsubscript{4}, ascorbic acid, AgNO\textsubscript{3}, and HCl all share the same 25 mL volume. To avoid confusion, the volume is explicitly associated with the first-mentioned precursor in the mixture, and the following precursors refer back to that first precursor. This ensures that downstream applications can unambiguously process the data to mean that the precursors are sharing a single volume. Other than these two errors, the model performs very well at extracting quantities in this example.

\begin{figure}[H]
\centering
\includegraphics[width=\linewidth]{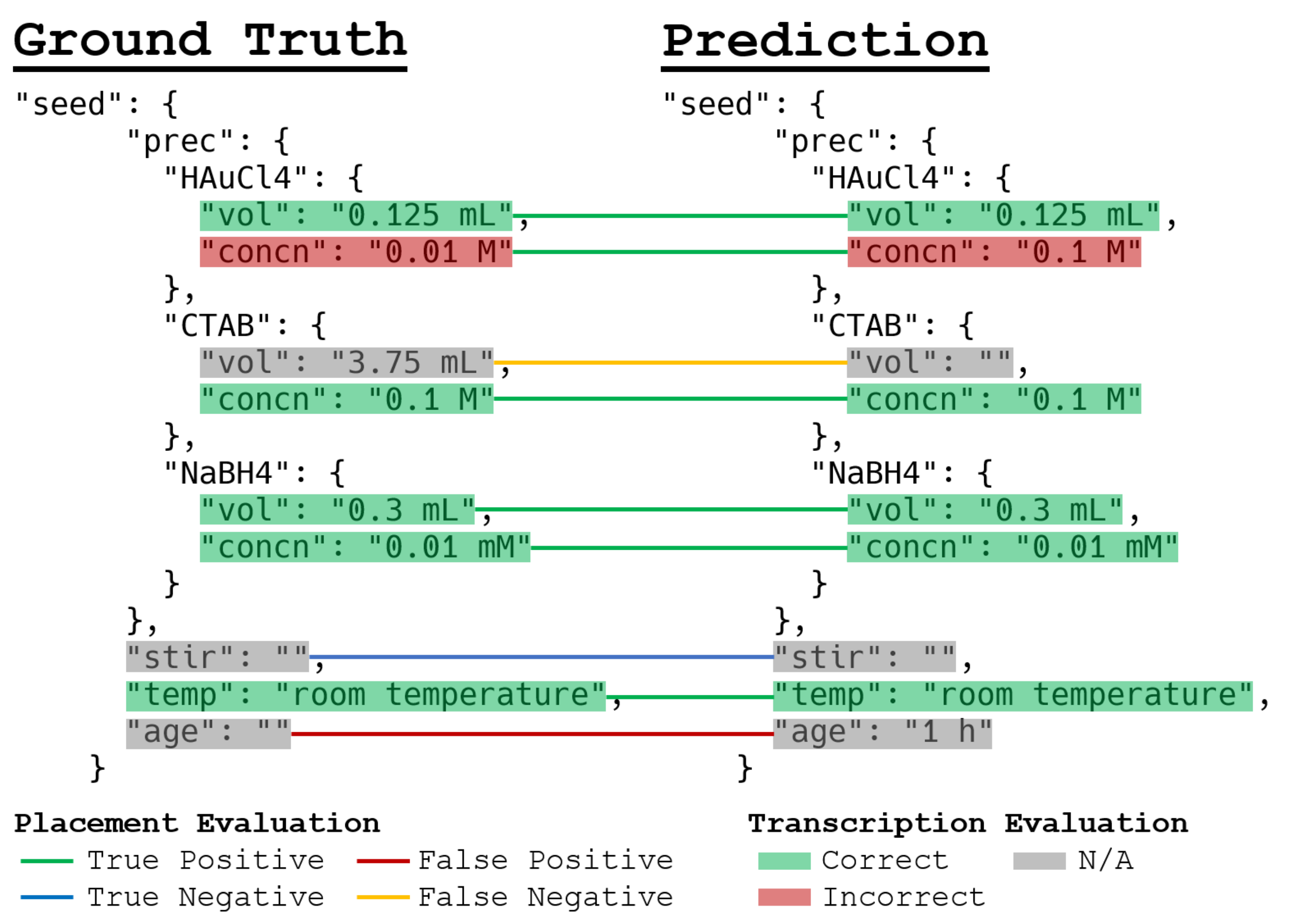}
\caption{A diagram depicting the different types of prediction errors made by the model is presented. Generally, two categories of errors exist: placement errors and transcription errors. Placement errors refer to whether the prediction has placed any information, correct or incorrect, into the appropriate fields as determined by the ground truth. These are indicated with the lines connecting the fields in the ground truth and the prediction templates. A false positive prediction occurs when the prediction places information in a field that is empty, while a false negative prediction is the reverse. A true negative prediction is when a field is empty in both the ground truth and the prediction, and a true positive prediction is when a field is non-empty in both the ground truth and the prediction. Since the placement evaluations do not consider whether the predicted value in a field is actually correct for true positives, an additional transcription evaluation is used to measure how well the predicted value explicitly matches the ground truth value. These are indicated with boxes encapsulating the fields. The transcription evaluation is only applied to true positive placements.}
\label{fig:evaluation}
\end{figure}

For the 117 testing paragraphs, two types of errors are tracked: placement errors and transcription errors. This is done in order to evaluate the model's capability for separately identifying which fields of the synthesis templates should contain information, as well as how accurate the appropriately placed information is. To evaluate information \textit{placement}, only the existence of information in the fields of the prediction and ground truth synthesis templates are considered. For example, if the same field contains information (as opposed to being empty) in both templates, that is considered a true positive prediction regardless of whether the information explicitly matches. If both fields are empty, then that is a true negative. If the prediction field contains information while the ground truth field is empty, then that is a false positive, while the reverse is a false negative. These categories of placement errors are used to calculate the precision, recall, and F1-score for information placement. Examples of these evaluations are shown in Figure \ref{fig:evaluation}.

For evaluating transcription accuracy, only the agreement between the prediction and the annotation for true positive placements are considered, as the other types of errors are accounted for by the evaluations of information placement. For numerical values with units, the units must be exactly correct and the quantitative relative error was calculated according to the function $s(p, q) = 2\cdot\min(p, q)/(p+q)$, which is derived from the absolute proportional difference $r(p, q) = \abs{p-q}/(p+q)$ and is bounded on $[0,1]$ for non-negative $p$ and $q$. Some values may have modifiers attached, such as ``$>3$ h''. If the prediction misses this information, \textit{e.g.}, gives ``3 h'', the prediction is considered half-correct even if the quantity and unit are both correct. Some quantities will additionally be expressed as a range or list of values. In these cases, the range boundaries are split into a list as necessary, and the transcription accuracies are scored and aggregated across the values in the list with proper ordering enforced. For non-numerical predictions such as stir rates described as ``vigorous'' or gold seed morphologies, an exact string match is required for the prediction to be marked as correct. The combined accuracy is presented as the product of the F1-score for information placement and the transcription accuracy.

\subsection{Model Performance}

\begin{table*}[!htbp]
\centering
\captionof{table}{Model F1-scores and accuracies for recipe entities aggregated by recipe component. The support numbers in parentheses account for only the true positives used for the accuracy calculation.}
\label{tab:aggregate_results_static}
\begin{tabular}{l|lll|l|l|l}
\cline{2-6}
 & \multicolumn{3}{c|}{Placement} & \multicolumn{1}{c|}{Transcription} & \multicolumn{1}{c|}{Combined} &  \\ \cline{2-7} 
 & \multicolumn{1}{l|}{Precision} & \multicolumn{1}{l|}{Recall} & F1 & Accuracy & Adj. F1 & \multicolumn{1}{l|}{Support} \\ \hline
\multicolumn{1}{|l|}{Seed Solution} & \multicolumn{1}{l|}{0.97} & \multicolumn{1}{l|}{0.92} & 0.94 & 0.95 & 0.90 & \multicolumn{1}{l|}{159 (142)} \\ \hline
\multicolumn{1}{|l|}{Growth Solution} & \multicolumn{1}{l|}{0.90} & \multicolumn{1}{l|}{0.94} & 0.92 & 0.96 & 0.88 & \multicolumn{1}{l|}{244 (206)} \\ \hline
\multicolumn{1}{|l|}{AuNR} & \multicolumn{1}{l|}{0.79} & \multicolumn{1}{l|}{0.74} & 0.76 & 0.95 & 0.72 & \multicolumn{1}{l|}{96 (59)} \\ \hline
\multicolumn{1}{|l|}{Overall} & \multicolumn{1}{l|}{\textbf{0.90}} & \multicolumn{1}{l|}{\textbf{0.90}} & \textbf{0.90} & \textbf{0.96} & \textbf{0.86} & \multicolumn{1}{l|}{499 (407)} \\ \hline
\end{tabular}
\end{table*}

The total performance of the model aggregated over each recipe component as well as all entries is shown in Table \ref{tab:aggregate_results_static}. The model appears to be proficient at generally identifying which information should be filled in the template based on the content of the text, with a rather high F1-score of $90\%$ that favors neither precision nor recall. It additionally performs exceptionally at accurately transcribing the information to provide an impressive overall score of $86\%$. This indicates a significant improvement over comparable efforts in solid-state synthesis text-mining, which report an overall accuracy of $51\%$ for extracting all recipe items (chemistry, operations, and attributes of the operations).\cite{matsynth_corpus} Direct comparison is, however, rather challenging, as some aspects of the two-step, seed-mediated growth synthesis are more complicated, such as the presence of two solutions with distinct precursor sets and a greater amount of precursor information needed due to the solution-based format. On the other hand, solid-state synthesis extraction carries its own challenges, considering the greater variation in procedural steps and conditions that must be considered.

It is clear that the F1-scores for the recipe entities associated with the seed and growth solutions are very promising, indicating that the model is reliable for extracting the necessary information from the text for the component solutions to the synthesis procedure. However, the performance is worse overall for the gold nanorod properties, with an adjusted F1-score of approximately $72\%$. This is still an improvement over established results, however, as the gold nanoparticle synthesis protocol and outcome database developed by Cruse et al.\cite{aunp_synth_dataset} extracts morphology measurements, sizes, and units with F1-scores of $70\%$, $69\%$, and $91\%$ via NER with MatBERT, \textit{without linking them together}. This would inevitably introduce additional sources of error and performance would be additionally constrained by the lowest performing extractions.

\begin{table*}[!htbp]
\centering
\caption{Model performance for precursor detection in the seed and growth solution information.}
\label{tab:solution_detection_results_static}
\begin{tabular}{l|llll|llll|}
\cline{2-9}
\multirow{2}{*}{} & \multicolumn{4}{l|}{Seed Solution} & \multicolumn{4}{l|}{Growth Solution} \\ \cline{2-9} 
 & \multicolumn{1}{l|}{Precision} & \multicolumn{1}{l|}{Recall} & \multicolumn{1}{l|}{F1} & Support & \multicolumn{1}{l|}{Precision} & \multicolumn{1}{l|}{Recall} & \multicolumn{1}{l|}{F1} & Support \\ \hline
\multicolumn{1}{|l|}{Precursor} & \multicolumn{1}{l|}{0.98} & \multicolumn{1}{l|}{0.90} & \multicolumn{1}{l|}{\textbf{0.94}} & 61 & \multicolumn{1}{l|}{0.93} & \multicolumn{1}{l|}{0.92} & \multicolumn{1}{l|}{\textbf{0.92}} & 118 \\ \hline
\end{tabular}
\end{table*}

Table \ref{tab:solution_detection_results_static} shows the model performance for detecting precursors in the seed and growth solutions. Precursor detection is calculated implicitly based on which precursors the extracted volumes, concentrations, and masses are associated with. This is a clear improvement over the results in the gold nanoparticle synthesis protocol and outcome database developed by Cruse et al.,\cite{aunp_synth_dataset}. The prior work detected precursors via a BiLSTM-based NER model with an F1-score of $90\%$ without distinguishing between seed or growth solution precursors and additionally without the capability for detecting precursors that do not contain specific formulae or chemical names, such as the seed solution that is added to the growth solution. The fine-tuned GPT-3 model missed cases where cationic surfactant, PP, BH\textsubscript{4}, and AuCl\textsubscript{3} were used as well as a case where HCl was used in the seed solution. None of these cases occurred in the training set. Notably, the model correctly normalized ``AsA'' to ``AA'', despite ``AsA'' never appearing in the training data.

\subsection{Full Dataset}

\begin{table*}[!htbp]
\centering
\caption{A table depicting the format of each data record for each article in the dataset is presented (constructed by merging paragraph templates). The ``doi'' key contains the article DOI and the ``text'' key contains index keys of the relevent paragraphs within that article which in turn contain the paragraph text. The ``seed'' and ``growth'' keys respectively contain the keys for the seed and growth solution information, including the ``prec'' key for precursors, the ``stir'' key for stir rates (when adding the reducing agent for the seed solution and when adding the seed solution for the growth solution), the ``temp'' key for the aging temperature, and the ``age'' key for the solution aging time. The ``seed'' key has an additional ``seed'' key that contains the ``size'' and ``shape'' keys for the size and shape of the seeds in the seed solution. The ``prec'' key for each solution contains multiple keys for each precursor in each solution, anonymized as ``<precursor name>'' in the table. For each precursor, there are three keys: ``vol'', ``concn'', and ``mass'' for the precursor volume, concentration, and mass, respectively. The ``AuNR'' key contains keys for measurements of gold nanorod dimensions: ``ar'', ``l'', ``w'', ``lspr'', and ``tspr'' for the aspect ratio, length, width, LSPR, and TSPR, respectively. Each extracted value is additionally stored as a key with a corresponding list of the paragraph indices that the value was extracted from in order to preserve information about entity sources. The final column displays the total number of entities extracted for each key (with no subkeys).}
\label{tab:document_structure}
\begin{tabular}{|l|l|l|l|p{4cm}|l|}
\hline
Root Key & First Subkey & Second Subkey & Third Subkey & Description & Total \\ \hline
doi &  &  &  & Article DOI & 1137 \\ \hline
text & \textless{}integer\textgreater{} &  &  & Paragraph text for \textless{}integer\textgreater{}\textsuperscript{th} paragraph & 2969 \\ \hline
\multirow{8}{*}{seed} & \multirow{3}{*}{prec} & \multirow{3}{*}{\textless{}precursor name\textgreater{}} & volume & Seed solution precursor volume & 1347 \\ \cline{4-6} 
 &  &  & concentration & Seed solution precursor concentration & 1385 \\ \cline{4-6} 
 &  &  & mass & Seed solution precursor mass & 6 \\ \cline{2-6} 
 & \multirow{2}{*}{seed} & size &  & Seed solution seed size & 137 \\ \cline{3-6} 
 &  & shape &  & Seed solution seed shape & 24 \\ \cline{2-6} 
 & stir &  &  & Seed solution reducing agent stir rate & 266 \\ \cline{2-6} 
 & temp &  &  & Seed solution aging temperature & 284 \\ \cline{2-6} 
 & age &  &  & Seed solution aging time & 352 \\ \hline
\multirow{6}{*}{growth} & \multirow{3}{*}{prec} & \multirow{3}{*}{\textless{}precursor name\textgreater{}} & volume & Growth solution precursor volume & 2664 \\ \cline{4-6} 
 &  &  & concentration & Growth solution precursor concentration & 2178 \\ \cline{4-6} 
 &  &  & mass & Growth solution precursor mass & 65 \\ \cline{2-6} 
 & stir &  &  & Growth solution reducing agent stir rate & 134 \\ \cline{2-6} 
 & temp &  &  & Growth solution aging temperature & 322 \\ \cline{2-6} 
 & age &  &  & Growth solution aging time & 464 \\ \hline
\multirow{5}{*}{AuNR} & ar &  &  & Gold nanorod aspect ratio & 587 \\ \cline{2-6} 
 & l &  &  & Gold nanorod length & 443 \\ \cline{2-6} 
 & w &  &  & Gold nanorod width & 452 \\ \cline{2-6} 
 & lspr &  &  & Gold nanorod LSPR & 357 \\ \cline{2-6} 
 & tspr &  &  & Gold nanorod TSPR & 177 \\ \hline
\end{tabular}
\end{table*}

The fine-tuned GPT-3 model was applied to the full dataset of 1,137 papers (2,969 paragraphs) at a total cost of 384.31 USD over 33 hours. In total, 11,644 entities were extracted from the paragraphs that contained information of interest. The dataset is presented as a JSON file containing a list with each element corresponding to a single article. Table \ref{tab:document_structure} summarizes the structure of the JSON documents for each paper alongside a breakdown of how the total extracted entities across the entire dataset are distributed across the entity types. While the template extractions were performed paragraph-by-paragraph, the templates have been merged by article for convenience. However, this does mean that some conflicts and repetitions are present in the dataset. A conflict arises when a particular entity type in a paper (e.g. the volume of a particular precursor) is specified with different values across multiple paragraphs and a repetition arises when it is specified with the same value across multiple paragraphs. Of the 11,644 extracted entities, 10,098 ($\sim 87\%$) are uniquely identified, meaning there are no conflicts or repetitions (the associated value is extracted from exactly one paragraph). An additional 353 entries present at least one conflict without any repetitions, 251 with at least one repetition and no conflicts, and 57 with both conflicts and repetitions. Repetitions do not need to be manually resolved since this arises from the specification of identical information across multiple paragraphs (e.g. mentioning the gold nanorod aspect ratios in paragraphs about both the synthesis procedure as well as the nanorod characterization), but conflicts can be challenging to resolve in a consistent manner without manual inspection. For instance, if two separate volumes for a particular precursor are provided in two separate paragraphs, it can be ambiguous whether the volumes are part of the same synthesis procedure or distinct synthesis procedures in the same paper due to the lack of cross-paragraph context. With this in mind, of the 11,644 extracted entities, 10,349 ($\sim 89\%$) can be safely extracted by automatically resolving repetitions and discarding entities with conflicts. Of the entities with conflicts, 341 have two distinct values, 47 have three, 12 have five, 9 have four, and 1 has five.

\begin{table*}[!htbp]
\centering
\caption{A table depicting the format of each extracted value in the post-processed version of the dataset.}
\label{tab:postproc_structure}
\begin{tabular}{|l|l|p{6cm}|}
\hline
Key & Structure & Description \\ \hline
mod & \textless{}modifier\textgreater{} & A string indicating if a value is a range, approximate, bounding, or unprocessed. \\ \hline
val & {[}\textless{}value\textgreater{}, ..., \textless{}value\textgreater{}{]} & A list of the extracted values. Ranges will consist of two values for the range boundaries. Processed values will be numbers while unprocessed values will be strings. \\ \hline
unit & \textless{}unit\textgreater{} & The units for the extracted values, if applicable, as a string. \\ \hline
src & {[}{[}\textless{}index\textgreater{}, ..., \textless{}index\textgreater{}{]}, ..., {[}...{]}{]} & A list of lists of paragraph indices to indicate the source for the extracted information. \\ \hline
index & {[}{[}\textless{}index\textgreater{}, ..., \textless{}index\textgreater{}{]}, ..., {[}...{]}{]} & A list of lists of positional indices to retain ordering for values that were split from a list during post-processing. \\ \hline
\end{tabular}
\end{table*}

With post-processing applied (as was done for evaluation of the testing dataset), splitting lists of extracted values into distinct entities and resolving repetitions of identical information extracted across different paragraphs within the same papers results in a total of 11,770 unique entities. In the post-processed version of the dataset, each property contains a list of dictionaries with structures indicated in Table \ref{tab:postproc_structure}.

\subsection{Full Dataset Analysis}

\subsubsection{Procedure Completeness Analysis}

\begin{figure}[H]
\centering
\includegraphics[width=\linewidth]{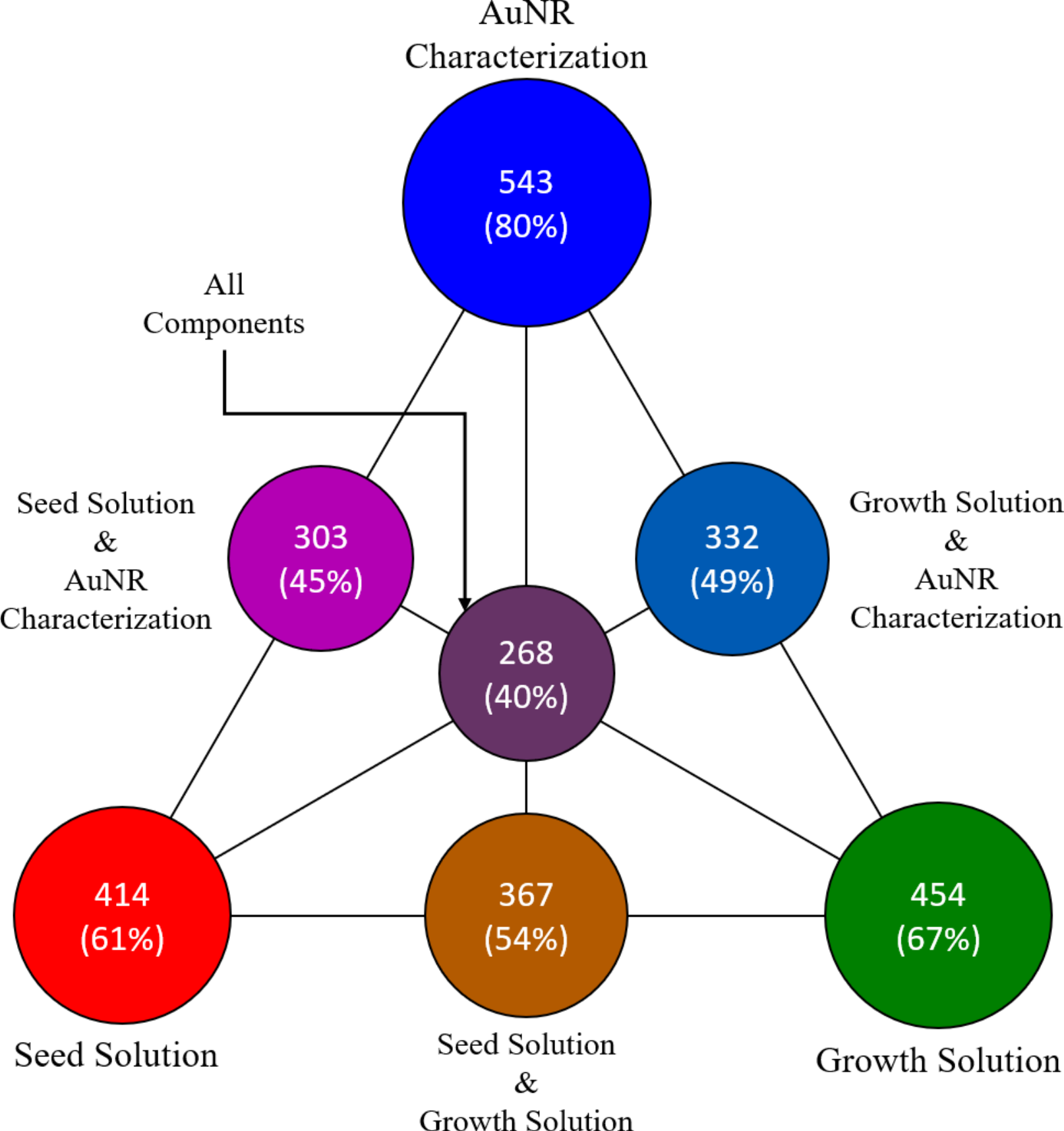}
\caption{A diagram showing the proportional overlaps of papers with complete synthesis procedure and outcome components. Each vertex of the triangle corresponds to the labeled recipe component. The areas of the circles are proportional to the corresponding number of papers inscribed. The circles on the midpoints of the edges correspond to papers with complete recipe components corresponding to the bounding vertices. The center circle corresponds to the papers with complete recipes and complete characterizations.}
\label{fig:triangle_full_relaxed}
\end{figure}

An ideal database of gold nanorod growth procedures should contain fully-specified, reproducible procedures alongside their outcomes. This is desirable because missing information could inhibit downstream applications that need complete information about the synthesis procedure. For instance, if a scientist wants to reproduce an experiment that produces gold nanorods of a particular aspect ratio, they would at the very least need to know all of the relevant seed and growth solution precursors with their amounts. Similarly, a data science project that intends to investigate the relationship between procedures and outcomes will need complete information for the seed and growth solutions in addition to the gold nanorod measurements in order to produce reliable predictions. To evaluate the completeness of the information this dataset contains, we examined 1,137 papers in the full prediction dataset. Of these, 701 (62\%) contained at least one paragraph with a non-empty synthesis template. Of these 701 papers, 678 (97\%) fully specified at least one synthesis component: the seed solution, the growth solution, or the gold nanorod dimensions.  This is encouraging since the vast majority of the papers that contain information at least fully specify one component of the procedure or the outcome.

In order to evaluate the completeness of the components of the procedure and the outcome, for seed and growth solutions, only fully specified precursors were considered necessary for reproducibility. Auxiliary information, such as stirring rates, aging times, aging temperatures, and seed particle morphologies and sizes, while useful, was not considered necessary. The precursor information was considered to be full specified for a given paper if all of the precursor quantities were fully specified with either volume and concentration, mass, or a specific concentration within another solution for each precursor with extracted quantities. Exceptions were made for water and the seed solution that is added to the growth solution, which both only needed a reported volume or mass. Additionally, seed solution in the growth solution precursors was required for the growth solution precursors to be considered complete. For the gold nanorod dimensions to be considered complete, either the aspect ratio, length, or LSPR measurement had to be specified, with the latter two at least providing an avenue for estimation of the aspect ratio if reported alone.

Figure \ref{fig:triangle_full_relaxed} shows how the papers in the full prediction dataset are distributed across fully-specified synthesis procedure and outcome components according to these criteria. The vast majority of the papers reported gold nanorod dimensions, with 80\% of the 678 papers with at least one fully specified synthesis component containing fully-specified gold nanorod dimensions. Additionally, the majority of the papers fully-specified the seed and growth solutions (respectively 61\% and 67\%). However, they are distributed such that 40\% (268) of the papers fully specified all three components. This is a reasonable result considering that many papers will directly report the relevant gold nanorod dimensions without specifying a synthesis procedure, opting instead to reference the established recipe that the researchers used to produce the gold nanorods. Additionally, some researchers will opt  to purchase gold seed solution instead of producing their own, which accounts for cases where some papers are missing information about seed solution preparation. Most of the papers with fully-specified synthesis procedures and outcomes (162) used the typical 8-precursor synthesis and an additional 49 use the same synthesis precursors with the addition of HCl in the growth solution. In the post-processed version of the dataset, it is determined that of the 268 papers that fully specified all three components, 233 contained exactly one procedure. An additional 16 contained two, 13 contained three, 3 contained four, 2 contained five, and 1 contained six for a total of 332 complete procedures. This final dataset should be suitable for downstream analysis and inference, given the overall model performance for extracting complete synthesis procedures and outcomes from the literature.

\subsubsection{Data Consistency Analysis}

\begin{figure}[H]
\centering
\includegraphics[width=\linewidth]{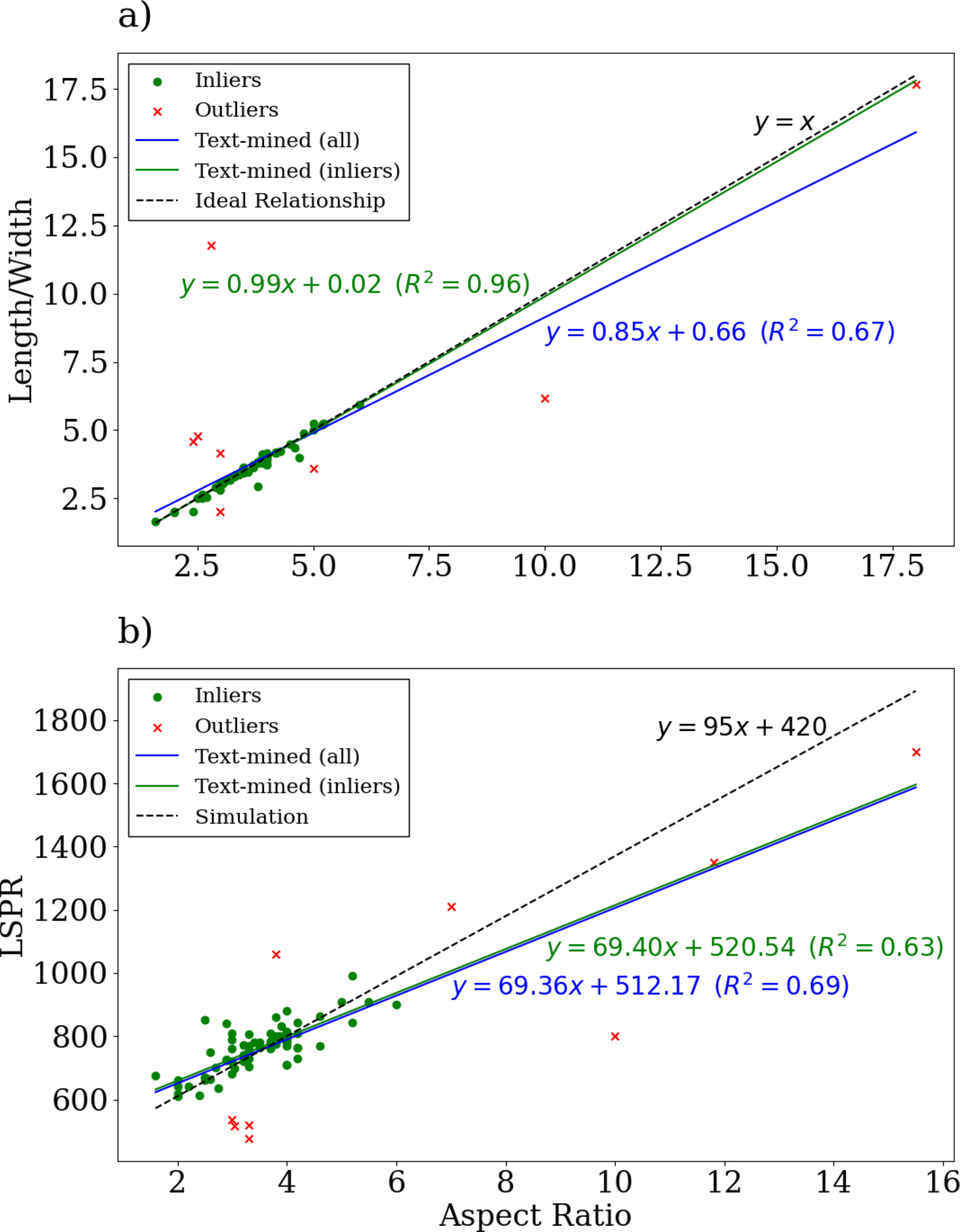}
\caption{A diagram showing the relationships between the gold nanorod aspect ratios and other gold nanorod measurements extracted from the literature including the (a) ratio between length and width and (b) the LSPR peak. The inlier datapoints are shown in purple and the outlier datapoints in red. The linear regressions derived from the text-mined data using all of the available data and only the inlier data are respectively shown in red and purple on each sub-diagram. For the comparison to the ratio between length and width (a), the ideal relation is shown with a dashed black line and for the LSPR comparison (b), a simulated relationship is shown with a dashed black line.\cite{ar_lspr_0}}
\label{fig:ar_relations}
\end{figure}

Figure \ref{fig:ar_relations} shows the relationship between various measurements extracted from text compared to the aspect ratios extracted from the text. Only co-occurring measurements explicitly present within the extracted information from a given paragraph are considered data points for comparison. No derived measurements were used. As a sanity check, the first diagram (a) shows the relationship between the ratios of the explicit lengths and widths present in the text (excluding ranges) and the reported aspect ratios. Ideally, the relationship should be an identity as shown with the dashed line. However, while the vast majority of the data approximately complies with this trend, there are several outliers that produce deviation from the ideal trend in the regression of the text-mined data. This is primarily caused by two papers with mismatches in measurements extracted from three-step seed-mediated gold nanorod overgrowth procedures where the dimensions of the nanorod seeds used for overgrowth into nanowires are confused with the dimensions of the nanowires themselves. With all outliers removed via outlier detection using an elliptic envelope followed by manual verification, the linear regression almost exactly matches the ideal relationship. The most common errors were caused by nanorod overgrowth measurements taken from three-step seed mediated growth procedures and cases in which the ordering of the aspect ratios and the lengths and widths were mismatched (e.g. the lengths and widths are listed while the aspect ratios are presented as a range). Only 8 of the 78 data points were identified as outliers. For the comparison between the LSPR peaks and the aspect ratios (b), a strong linear trend is similarly present. However, for this relationship, there is an additional comparison to a relationship derived from simulation using a set refractive index for gold nanorods shown in blue, which is in general agreement with the relationship derived from text-mined empirical data.\cite{ar_lspr_0} The deviations can be explained by multiple factors including deviations from ideal conditions shifting the LSPR peaks such as deviation from spherical end-cap geometries, low nanorod yields, or impurities in the gold nanorod solution or the nanorods themselves that change the refractive index (including poor cleaning or high concentrations of silver in procedures using AgNO\textsubscript{3}).\cite{ar_lspr_0,ar_lspr_1} While there are extraction errors present, outlier removal using an elliptic envelope followed by manual verification does not significantly change the linear regression. Outliers were most commonly caused by extraction errors that swapped the LSPR and TSPR measurements provided in the text. Only 9 of the 86 data points were identified as outliers. Deviation from the theory in such a manner is to be expected when considering empirical data from real-world experiments. Still, the LSPR for spheres should be around 520 nm while the text-mined trend line points towards a value closer to $580\sim 590$ nm. However, for larger aspect ratios, the text-mined trend line is more representative of the text-mined empirical data than the trend line derived from simulation. The major outlier present in the text-mined data is once again explained by a mismatch in measurements from a three-step seed-mediated gold nanorod overgrowth procedure.

\subsubsection{Gold Nanorod Aspect Ratio Distribution Analysis}

\begin{figure}[H]
\centering
\includegraphics[width=\linewidth]{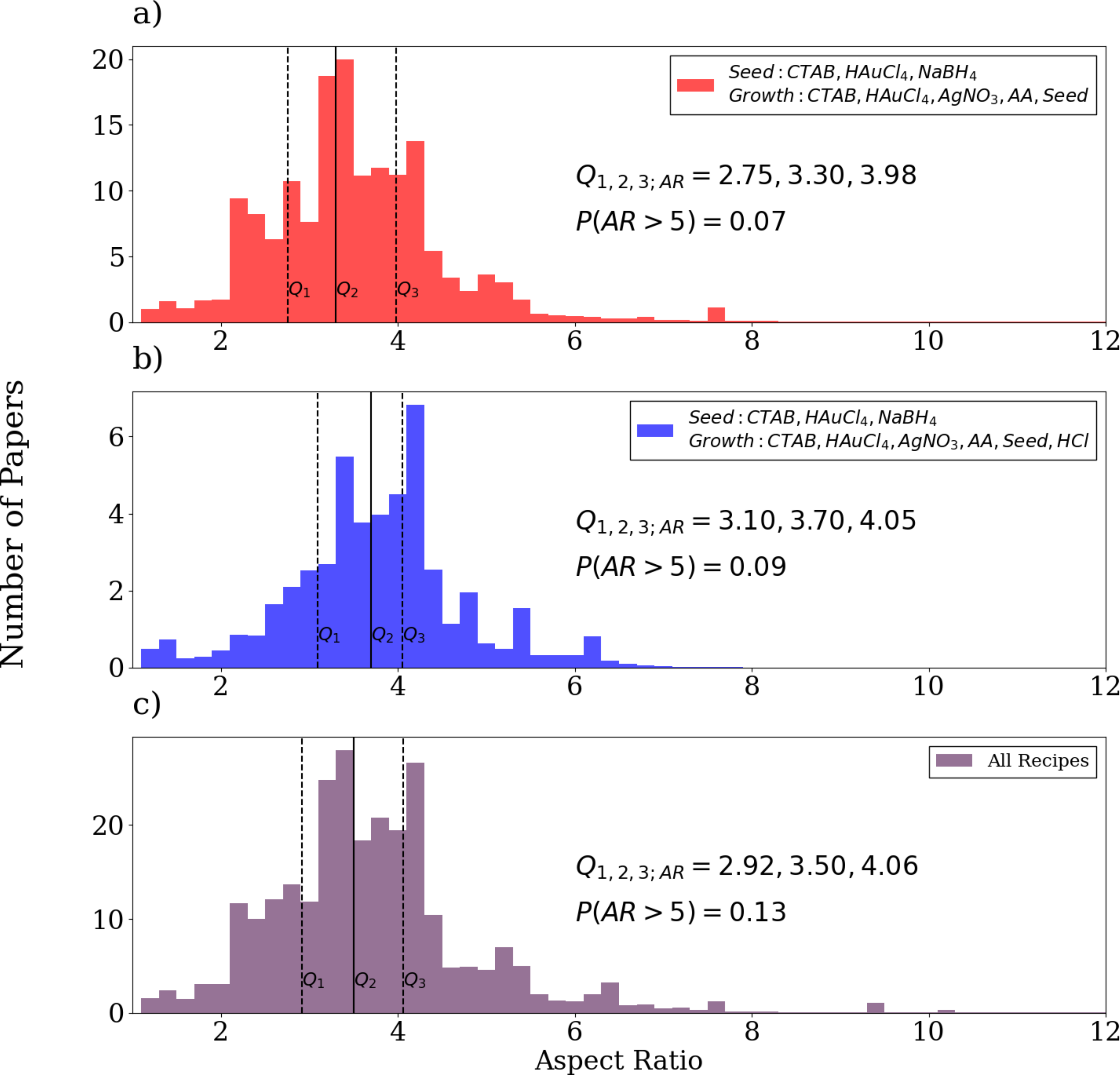}
\caption{A diagram showing the distributions of gold nanorod aspect ratios resulting from different precursor sets including the (a) standard procedure, (b) the addition of HCl in the growth solution, and (c) all complete precursor sets. Negligible contributions for aspect ratios larger than 20 are not shown ($P(AR > 12) < 0.02$). In each sub-diagram, the median is shown with a solid black line and the first and third quartiles are shown with dashed black lines.}
\label{fig:ar_distribution}
\end{figure}

Figure \ref{fig:ar_distribution} shows the distributions of the aspect ratios extracted from fully-specified experiments using precursor sets found in more than 10 papers in the full prediction database (Figures \ref{fig:ar_distribution}a and \ref{fig:ar_distribution}b), in addition to the complete set of papers (Figure \ref{fig:ar_distribution}c). For many of the papers, the aspect ratios were directly reported. However, there are multiple different ways that they are reported that must be addressed in order to properly construct the distributions. If the aspect ratio is provided as a range of values, the distribution across that range was taken to be a normal distribution with a mean and standard deviation determined by the midpoint and endpoints of the range, respectively. For papers that did not report aspect ratios directly, length and width information was used instead. In cases where the lengths and widths were presented as ranges, they were similarly cast as normal distributions, and the distributions of the aspect ratios were calculated as ratio distributions. For cases where only the LSPR was provided, the text-mined linear relationship with outliers removed shown in Figure \ref{fig:ar_relations} was used to estimate the aspect ratios. In cases where any quantities were accompanied by an approximation modifier (e.g. $\sim$), the values were cast as uniform distributions over the range of $\pm 10\%$ of the value. Any calculated aspect ratios that fell below 1 (e.g. due to overlaps in length and width distributions for gold nanorods with small aspect ratios) were inverted.

From the distribution of the standard recipe, it is readily apparent that the median nanorod aspect ratio is $3.3$ with respective first and third quartiles of $2.75$ and $3.98$. Comparing with experiments reporting that varying the concentration of AgNO\textsubscript{3} in the growth solution varies the resulting nanorod aspect ratios from $1.83$ to $5.04$,\cite{ar_range} the distribution of gold nanorod aspect ratios text-mined from the literature is consistent with this range, though it is narrower. Notably, there is a non-negligible amount of samples with aspect ratios greater than 5 in the distribution for the standard procedure. This is not consistent with heuristic knowledge of the limitations of the standard procedure for producing large aspect ratio gold nanorods, usually due to shorter growth times compared to procedures that adjust the pH of the growth solution to retard the nanorod growth.\cite{growth_limits,limited_growth} This is primarily due to erroneous extractions of nanowire measurements from overgrowth experiments or missed precursors based on manual inspection of the data. However, the statistics are still dominated by the lower aspect ratios. Comparing to the distribution for experiments using HCl in the growth solution, it is apparent that the addition produces a distribution shifted towards larger aspect ratios. This is consistent with experiments that have determined that the use of HCl in the growth solution grants broader tunability of the gold nanorod aspect ratios, allowing for more controlled growth of longer nanorods relative to the standard procedure.\cite{hcl_growth_0,hcl_growth_1} Notably, $\sim 7\%$ of the procedures using the standard procedure and $\sim 9\%$ of the procedures using HCl in the growth solution provide nanorods with aspect ratios of 5 or higher. However, when all recipes are considered, it is clear that even longer nanorods can be synthesized, though these recipes are not as popular in the literature.

\section{Discussion}\label{discussion}

Overall, the model performs well at identifying and extracting relevant information specific to seed-mediated gold nanorod growth procedures in the literature. The model achieves an overall adjusted F1-score of $86\%$ on the testing dataset, indicating that it performs rather well at the task of simultaneous entity recognition and relation extraction. However, due to the static nature of the relations provided by the synthesis template and the single inference step, the entity recognition and relation extraction tasks are not easily disentangled, which limits direct comparison with conventional two-step approaches. Instead, the model performance for information retrieval is evaluated according to its ability to place information into fields of the template where information should exist and then the accuracy of the information that is correctly placed. For information placement, the precision, recall, and F1-score are balanced at $90\%$, indicating notable performance with no preference for false positives or false negatives. Of the information that is correctly placed in the templates, the model predicts the specific values with $96\%$ accuracy. Thus, the primary source of error is the accurate placement of information into the template rather than the accurate prediction of correctly placed information. However, the template model struggles with identifying new precursors that were not present in the training set.

The dataset produced by the model provides a wealth of information about seed-mediated gold nanorod growth experiments and, to our knowledge, constitutes the largest structured database with this level of depth and completeness. The model's ability to distinguish between precursors in the seed and growth solutions provides an example of very useful information. The simultaneous identification of precursors alongside linking them to the appropriate solutions in the two-step seed-mediated procedure had proven difficult using established methods due to the propagation of errors introduced by the reliance on separate models for entity extraction and relation. However, with this model, if a researcher wants to quickly find papers that used a particular precursor in the seed solution for seed-mediated growth of gold nanorods, this task can be accomplished with high fidelity using the predicted templates. Access to this information can be expected to greatly improve tools for scientific literature searches, as conventional simple keyword searches do not offer this specific relational dependence for complicated multi-step procedures.

For a more ambitious goal, the full synthesis procedure data can be leveraged for multiple downstream tasks, which would require the creation of additional models for inference. One example would be a model that predicts gold nanorod dimensions conditioned on a specific synthesis procedure: $p\pqty{\textrm{properties}|\textrm{procedure}}$. Such a model may be leveraged to predict the outcomes of proposed procedures without the need to perform them explicitly. Building on this, the inverse problem, $p\pqty{\textrm{procedure}|\textrm{properties}}$, can also be modeled. This would be very useful for streamlining synthesis experiments, as the necessary procedures for synthesizing gold nanorods with the desired properties can be inferred to provide a starting point that reduces the number of experiments that must be conducted to synthesize the desired gold nanorods. However, in the most likely case, any model trained on literature data alone will be incomplete and require further data generation and fine tuning.

Furthermore, it is worth considering how these templates fit into a larger project for downstream synthesis outcome predictions and synthesis procedure recommendations. The data extracted from literature can be used to pre-train models used for these purposes, while explicit experimental data can be used to further train the models to produce better predictions. The new templates provided by the experimental results are expected to be of extremely high quality, which will mitigate the errors present in the pre-training data from literature over time as more experimental results are added to the template database.

While this dataset is restricted to seed-mediated gold nanorod growth, the flexibility and performance of the templating approach using GPT-3 motivates application to other tasks for structured information retrieval from unstructured scientific text as has been shown in recent literature.\cite{gpt3_mastci} To this end, the dataset can be extended to accommodate seed-mediated growth of other gold nanoparticle morphologies, which may even improve overall model performance, as many errors were caused by the model erroneously extracting information from procedures that mentioned nanorod morphologies, but synthesized a different morphology. Additionally, more complex synthesis methods, such as three-step processes in which nanorods are first synthesized via seed-mediated growth to be used as seeds in a growth solution for overgrowth into nanowires, as well as other synthesis methods, such as citrate reduction, may require the creation of new templates and fine-tuning a separate model for each synthesis method to improve overall performance. Generally, it can be expected that more complex templates will require more examples for fine-tuning.

\section{Conclusions}\label{conclusions}

The presented model for static structured templating of seed-mediated gold nanorod growth procedures extracted from unstructured text using GPT-3 is demonstrated to be a promising approach for constructing high-quality structured databases of information from the scientific literature. This approach for extracting seed-mediated gold nanorod procedures and outcomes achieves an impressive adjusted F1-score of $86\%$ for the simultaneous identification and linking of synthesis procedure components. We present a final dataset of 11,644 entities extracted from 1,137 papers, resulting in 268 papers with at least one complete seed-mediated gold nanorod growth procedure and outcome for a total of 332 complete procedures. This method can potentially be utilized for many downstream applications including procedure searches oriented around specific features, statistical analysis of synthesis outcomes, synthesis outcome predictions conditioned on procedures, and synthesis procedure recommendations conditioned on outcomes among others given the wealth of structured information present. Overall, we present this approach as a flexible candidate for general-purpose structured data extraction from unstructured scientific text and contribute a dataset that may serve as a useful tool for investigating synthesis pathways beyond heuristics.

\section*{Data and Code Availability}
The data composed of DOIs and associated structured JSON outputs can be found online at \url{https://doi.org/10.6084/m9.figshare.19719310.v3}.\cite{gpt3_data} The texts for the paragraphs in each paper are excerpted due to copyright restrictions.

\section*{Author Contributions}
A.J., G.C., and K.A.P. supervised the research. K.C. wrote the data collection infrastructure, performed the data collection, and wrote and applied the initial gold nanoparticle article classification and information extraction models. S.G. provided experimental domain knowledge necessary for the template design. J.D. introduced the GPT-3 sequence-to-sequence information extraction methodology and prepared the graphic representation of the extraction template. N.W. co-developed the GPT-3 sequence-to-sequence information extraction methodology, designed the extraction templates, wrote the code for interfacing with GPT-3, performed all annotations, performed all GPT-3 experiments, and prepared all results. S.L. provided additional result validation. All authors contributed to the discussion and writing of the manuscript.

\section*{Conflicts of interest}
There are no conflicts to declare.

\section*{Acknowledgements}
This work was funded and intellectually led by the U.S. Department of Energy, Office of Science, Office of Basic Energy Sciences, Materials Sciences and Engineering Division under Contract No. DE-AC02-05CH11231 (D2S2 program KCD2S2). Additional funding used for data set generation via the OpenAI API was provided by Toyota Research Institute through the Accelerated Materials Design and Discovery program. This research used resources of the National Energy Research Scientific Computing Center (NERSC), a U.S. Department of Energy Office of Science User Facility operated under Contract No. DE-AC02-05CH11231. This work also used the Extreme Science and Engineering Discovery Environment (XSEDE) GPU resources, specifically the Bridges-2 supercomputer at the Pittsburgh Supercomputing Center, through allocation TG-DMR970008S.\cite{xsede} 

\insertbibliography{main}


\begin{thebibliography}{10}
\providecommand{\url}[1]{#1}
\csname url@samestyle\endcsname
\providecommand{\newblock}{\relax}
\providecommand{\bibinfo}[2]{#2}
\providecommand{\BIBentrySTDinterwordspacing}{\spaceskip=0pt\relax}
\providecommand{\BIBentryALTinterwordstretchfactor}{4}
\providecommand{\BIBentryALTinterwordspacing}{\spaceskip=\fontdimen2\font plus
\BIBentryALTinterwordstretchfactor\fontdimen3\font minus
  \fontdimen4\font\relax}
\providecommand{\BIBforeignlanguage}[2]{{%
\expandafter\ifx\csname l@#1\endcsname\relax
\typeout{** WARNING: IEEEtran.bst: No hyphenation pattern has been}%
\typeout{** loaded for the language `#1'. Using the pattern for}%
\typeout{** the default language instead.}%
\else
\language=\csname l@#1\endcsname
\fi
#2}}
\providecommand{\BIBdecl}{\relax}
\BIBdecl

\bibitem{aunp_hist_0}
\BIBentryALTinterwordspacing
S.~{Mohan Bhagyaraj} and O.~S. Oluwafemi, ``Chapter 1 - nanotechnology: The
  science of the invisible,'' in \emph{Synthesis of Inorganic Nanomaterials},
  ser. Micro and Nano Technologies, S.~{Mohan Bhagyaraj}, O.~S. Oluwafemi,
  N.~Kalarikkal, and S.~Thomas, Eds.\hskip 1em plus 0.5em minus 0.4em\relax
  Woodhead Publishing, 2018, pp. 1--18. [Online]. Available:
  \url{https://www.sciencedirect.com/science/article/pii/B9780081019757000014}
\BIBentrySTDinterwordspacing

\bibitem{aunp_hist_1}
P.~Colomban, M.~Gironda, G.~Simsek~Franci, and P.~d'Abrigeon,
  ``\BIBforeignlanguage{en}{Distinguishing genuine imperial qing dynasty
  porcelain from ancient replicas by {On-Site} {Non-Invasive} {XRF} and raman
  spectroscopy},'' \emph{\BIBforeignlanguage{en}{Materials (Basel)}}, vol.~15,
  no.~16, Aug. 2022.

\bibitem{aunp_hist_2}
\BIBentryALTinterwordspacing
S.~Szunerits and R.~Boukherroub, ``Near-infrared photothermal heating with gold
  nanostructures,'' in \emph{Encyclopedia of Interfacial Chemistry},
  K.~Wandelt, Ed.\hskip 1em plus 0.5em minus 0.4em\relax Oxford: Elsevier,
  2018, pp. 500--510. [Online]. Available:
  \url{https://www.sciencedirect.com/science/article/pii/B9780124095472132287}
\BIBentrySTDinterwordspacing

\bibitem{aunr_hist}
\BIBentryALTinterwordspacing
S.~E. Lohse and C.~J. Murphy, ``The quest for shape control: A history of gold
  nanorod synthesis,'' \emph{Chemistry of Materials}, vol.~25, no.~8, pp.
  1250--1261, Apr 2013. [Online]. Available:
  \url{https://doi.org/10.1021/cm303708p}
\BIBentrySTDinterwordspacing

\bibitem{aunr_synth}
\BIBentryALTinterwordspacing
N.~D. Burrows, S.~Harvey, F.~A. Idesis, and C.~J. Murphy, ``Understanding the
  seed-mediated growth of gold nanorods through a fractional factorial design
  of experiments,'' \emph{Langmuir}, vol.~33, no.~8, pp. 1891--1907, Feb 2017.
  [Online]. Available: \url{https://doi.org/10.1021/acs.langmuir.6b03606}
\BIBentrySTDinterwordspacing

\bibitem{aunr_synth_2}
\BIBentryALTinterwordspacing
L.~Gou and C.~J. Murphy, ``Fine-tuning the shape of gold nanorods,''
  \emph{Chemistry of Materials}, vol.~17, no.~14, pp. 3668--3672, Jul 2005.
  [Online]. Available: \url{https://doi.org/10.1021/cm050525w}
\BIBentrySTDinterwordspacing

\bibitem{aunr_optical_0}
\BIBentryALTinterwordspacing
P.~K. Jain, X.~Huang, I.~H. El-Sayed, and M.~A. El-Sayed, ``Noble metals on the
  nanoscale: Optical and photothermal properties and some applications in
  imaging, sensing, biology, and medicine,'' \emph{Accounts of Chemical
  Research}, vol.~41, no.~12, pp. 1578--1586, Dec 2008. [Online]. Available:
  \url{https://doi.org/10.1021/ar7002804}
\BIBentrySTDinterwordspacing

\bibitem{aunr_optical_1}
E.~C. Dreaden, A.~M. Alkilany, X.~Huang, C.~J. Murphy, and M.~A. El-Sayed,
  ``\BIBforeignlanguage{en}{The golden age: gold nanoparticles for
  biomedicine},'' \emph{\BIBforeignlanguage{en}{Chem Soc Rev}}, vol.~41, no.~7,
  pp. 2740--2779, Nov. 2011.

\bibitem{aunr_optical_2}
S.~Eustis and M.~A. El-Sayed, ``Why gold nanoparticles are more precious than
  pretty gold: noble metal surface plasmon resonance and its enhancement of the
  radiative and nonradiative properties of nanocrystals of different shapes,''
  \emph{Chemical society reviews}, vol.~35, no.~3, pp. 209--217, 2006.

\bibitem{aunr_optical_3}
J.~C. Hulteen and C.~R. Martin, ``A general template-based method for the
  preparation of nanomaterials,'' \emph{Journal of materials chemistry},
  vol.~7, no.~7, pp. 1075--1087, 1997.

\bibitem{aunp_semiconductor}
\BIBentryALTinterwordspacing
K.~Sandeep, B.~Manoj, and K.~G. Thomas, ``Gold nanoparticle on semiconductor
  quantum dot: Do surface ligands influence fermi level equilibration,''
  \emph{The Journal of Chemical Physics}, vol. 152, no.~4, p. 044710, 2020.
  [Online]. Available: \url{https://doi.org/10.1063/1.5138216}
\BIBentrySTDinterwordspacing

\bibitem{aunp_semiconductor_2}
\BIBentryALTinterwordspacing
M.~Lau, A.~Ziefuss, T.~Komossa, and S.~Barcikowski, ``Inclusion of supported
  gold nanoparticles into their semiconductor support,'' \emph{Phys. Chem.
  Chem. Phys.}, vol.~17, pp. 29\,311--29\,318, 2015. [Online]. Available:
  \url{http://dx.doi.org/10.1039/C5CP04296H}
\BIBentrySTDinterwordspacing

\bibitem{aunp_biomedical}
\BIBentryALTinterwordspacing
L.~A. Dykman and N.~G. Khlebtsov, ``\BIBforeignlanguage{eng}{Gold nanoparticles
  in biology and medicine: recent advances and prospects},''
  \emph{\BIBforeignlanguage{eng}{Acta naturae}}, vol.~3, no.~2, pp. 34--55, Apr
  2011, 22649683[pmid]. [Online]. Available:
  \url{https://pubmed.ncbi.nlm.nih.gov/22649683}
\BIBentrySTDinterwordspacing

\bibitem{aunp_biomedical_2}
\BIBentryALTinterwordspacing
X.~Huang and M.~A. El-Sayed, ``Gold nanoparticles: Optical properties and
  implementations in cancer diagnosis and photothermal therapy,'' \emph{Journal
  of Advanced Research}, vol.~1, no.~1, pp. 13--28, 2010. [Online]. Available:
  \url{https://www.sciencedirect.com/science/article/pii/S2090123210000056}
\BIBentrySTDinterwordspacing

\bibitem{aunp_cosmetics}
\BIBentryALTinterwordspacing
S.~Kaul, N.~Gulati, D.~Verma, S.~Mukherjee, and U.~Nagaich,
  ``\BIBforeignlanguage{eng}{Role of nanotechnology in cosmeceuticals: A review
  of recent advances},'' \emph{\BIBforeignlanguage{eng}{Journal of
  pharmaceutics}}, vol. 2018, pp. 3\,420\,204--3\,420\,204, Mar 2018,
  29785318[pmid]. [Online]. Available:
  \url{https://pubmed.ncbi.nlm.nih.gov/29785318}
\BIBentrySTDinterwordspacing

\bibitem{aunp_size_morph}
\BIBentryALTinterwordspacing
K.~I. Requejo, A.~V. Liopo, P.~J. Derry, and E.~R. Zubarev, ``Accelerating gold
  nanorod synthesis with nanomolar concentrations of poly(vinylpyrrolidone),''
  \emph{Langmuir}, vol.~33, no.~44, pp. 12\,681--12\,688, Nov 2017. [Online].
  Available: \url{https://doi.org/10.1021/acs.langmuir.7b02942}
\BIBentrySTDinterwordspacing

\bibitem{aunp_size}
\BIBentryALTinterwordspacing
Y.~C. Dong, M.~Hajfathalian, P.~S.~N. Maidment, J.~C. Hsu, P.~C. Naha,
  S.~Si-Mohamed, M.~Breuilly, J.~Kim, P.~Chhour, P.~Douek, H.~I. Litt, and
  D.~P. Cormode, ``Effect of gold nanoparticle size on their properties as
  contrast agents for computed tomography,'' \emph{Scientific Reports}, vol.~9,
  no.~1, p. 14912, Oct 2019. [Online]. Available:
  \url{https://doi.org/10.1038/s41598-019-50332-8}
\BIBentrySTDinterwordspacing

\bibitem{aunp_size_shape}
\BIBentryALTinterwordspacing
S.~A. Ng, K.~A. Razak, A.~A. Aziz, and K.~Y. Cheong, ``The effect of size and
  shape of gold nanoparticles on thin film properties,'' \emph{Journal of
  Experimental Nanoscience}, vol.~9, no.~1, pp. 64--77, 2014. [Online].
  Available: \url{https://doi.org/10.1080/17458080.2013.813651}
\BIBentrySTDinterwordspacing

\bibitem{aunp_syst}
\BIBentryALTinterwordspacing
C.~{Daruich De Souza}, B.~{Ribeiro Nogueira}, and M.~E.~C. Rostelato, ``Review
  of the methodologies used in the synthesis gold nanoparticles by chemical
  reduction,'' \emph{Journal of Alloys and Compounds}, vol. 798, pp. 714--740,
  2019. [Online]. Available:
  \url{https://www.sciencedirect.com/science/article/pii/S092583881931833X}
\BIBentrySTDinterwordspacing

\bibitem{aunp_form_model}
\BIBentryALTinterwordspacing
E.~Agunloye, L.~Panariello, A.~Gavriilidis, and L.~Mazzei, ``A model for the
  formation of gold nanoparticles in the citrate synthesis method,''
  \emph{Chemical Engineering Science}, vol. 191, pp. 318--331, 2018. [Online].
  Available:
  \url{https://www.sciencedirect.com/science/article/pii/S0009250918304160}
\BIBentrySTDinterwordspacing

\bibitem{aunp_shape_control_0}
\BIBentryALTinterwordspacing
M.~L. Personick and C.~A. Mirkin, ``Making sense of the mayhem behind shape
  control in the synthesis of gold nanoparticles,'' \emph{Journal of the
  American Chemical Society}, vol. 135, no.~49, pp. 18\,238--18\,247, Dec 2013.
  [Online]. Available: \url{https://doi.org/10.1021/ja408645b}
\BIBentrySTDinterwordspacing

\bibitem{aunp_shape_control_1}
M.~Grzelczak, J.~P{\'e}rez-Juste, P.~Mulvaney, and L.~M. Liz-Marz{\'a}n,
  ``Shape control in gold nanoparticle synthesis,'' \emph{Colloidal Synthesis
  of Plasmonic Nanometals}, pp. 197--220, 2020.

\bibitem{aunp_dft}
\BIBentryALTinterwordspacing
D.~F. Mukhamedzyanova, N.~K. Ratmanova, D.~A. Pichugina, and N.~E. Kuz'menko,
  ``A structural and stability evaluation of au12 from an isolated cluster to
  the deposited material,'' \emph{The Journal of Physical Chemistry C}, vol.
  116, no.~21, pp. 11\,507--11\,518, May 2012. [Online]. Available:
  \url{https://doi.org/10.1021/jp212367z}
\BIBentrySTDinterwordspacing

\bibitem{aunp_ligand_0}
\BIBentryALTinterwordspacing
M.~Domingo, M.~Shahrokhi, I.~N. Remediakis, and N.~Lopez, ``Shape control in
  gold nanoparticles by n-containing ligands: Insights from density functional
  theory and wulff constructions,'' \emph{Topics in Catalysis}, vol.~61, no.~5,
  pp. 412--418, May 2018. [Online]. Available:
  \url{https://doi.org/10.1007/s11244-017-0880-3}
\BIBentrySTDinterwordspacing

\bibitem{aunp_ligand_1}
\BIBentryALTinterwordspacing
I.~Chakraborty and T.~Pradeep, ``Atomically precise clusters of noble metals:
  Emerging link between atoms and nanoparticles,'' \emph{Chemical Reviews},
  vol. 117, no.~12, pp. 8208--8271, Jun 2017. [Online]. Available:
  \url{https://doi.org/10.1021/acs.chemrev.6b00769}
\BIBentrySTDinterwordspacing

\bibitem{aunp_continuum}
\BIBentryALTinterwordspacing
D.~V. Talapin, A.~L. Rogach, M.~Haase, and H.~Weller, ``Evolution of an
  ensemble of nanoparticles in a colloidal solution:{\thinspace} theoretical
  study,'' \emph{The Journal of Physical Chemistry B}, vol. 105, no.~49, pp.
  12\,278--12\,285, Dec 2001. [Online]. Available:
  \url{https://doi.org/10.1021/jp012229m}
\BIBentrySTDinterwordspacing

\bibitem{nlp_matsci_mining_1}
O.~Kononova, T.~He, H.~Huo, A.~Trewartha, E.~A. Olivetti, and G.~Ceder,
  ``{{O}pportunities and challenges of text mining in materials research},''
  \emph{iScience}, vol.~24, no.~3, p. 102155, Mar 2021.

\bibitem{matsynth_corpus}
\BIBentryALTinterwordspacing
O.~Kononova, H.~Huo, T.~He, Z.~Rong, T.~Botari, W.~Sun, V.~Tshitoyan, and
  G.~Ceder, ``Text-mined dataset of inorganic materials synthesis recipes,''
  \emph{Scientific Data}, vol.~6, no.~1, p. 203, Oct 2019. [Online]. Available:
  \url{https://doi.org/10.1038/s41597-019-0224-1}
\BIBentrySTDinterwordspacing

\bibitem{ner_chemistry_2}
\BIBentryALTinterwordspacing
S.~Eltyeb and N.~Salim, ``\BIBforeignlanguage{eng}{Chemical named entities
  recognition: a review on approaches and applications},''
  \emph{\BIBforeignlanguage{eng}{Journal of cheminformatics}}, vol.~6, pp.
  17--17, Apr 2014. [Online]. Available:
  \url{https://doi.org/10.1186/1758-2946-6-17}
\BIBentrySTDinterwordspacing

\bibitem{ner_chemistry_3}
\BIBentryALTinterwordspacing
P.~Corbett and J.~Boyle, ``Chemlistem: chemical named entity recognition using
  recurrent neural networks,'' \emph{Journal of Cheminformatics}, vol.~10,
  no.~1, p.~59, Dec 2018. [Online]. Available:
  \url{https://doi.org/10.1186/s13321-018-0313-8}
\BIBentrySTDinterwordspacing

\bibitem{ner_medical}
\BIBentryALTinterwordspacing
Z.~Liang, J.~Chen, Z.~Xu, Y.~Chen, and T.~Hao, ``A pattern-based method for
  medical entity recognition from chinese diagnostic imaging text,''
  \emph{Frontiers in Artificial Intelligence}, vol.~2, p.~1, 2019. [Online].
  Available: \url{https://www.frontiersin.org/article/10.3389/frai.2019.00001}
\BIBentrySTDinterwordspacing

\bibitem{ner_biomedical}
\BIBentryALTinterwordspacing
A.~Sniegula, A.~Poniszewska-Maranda, and L.~Chomatek, ``Study of named entity
  recognition methods in biomedical field,'' \emph{Procedia Computer Science},
  vol. 160, pp. 260--265, 2019, the 10th International Conference on Emerging
  Ubiquitous Systems and Pervasive Networks (EUSPN-2019) / The 9th
  International Conference on Current and Future Trends of Information and
  Communication Technologies in Healthcare (ICTH-2019) / Affiliated Workshops.
  [Online]. Available:
  \url{https://www.sciencedirect.com/science/article/pii/S1877050919316813}
\BIBentrySTDinterwordspacing

\bibitem{bioelectra}
\BIBentryALTinterwordspacing
K.~r. Kanakarajan, B.~Kundumani, and M.~Sankarasubbu,
  ``{B}io{ELECTRA}:pretrained biomedical text encoder using discriminators,''
  in \emph{Proceedings of the 20th Workshop on Biomedical Language
  Processing}.\hskip 1em plus 0.5em minus 0.4em\relax Online: Association for
  Computational Linguistics, Jun. 2021, pp. 143--154. [Online]. Available:
  \url{https://aclanthology.org/2021.bionlp-1.16}
\BIBentrySTDinterwordspacing

\bibitem{weston_ner}
L.~Weston, V.~Tshitoyan, J.~Dagdelen, O.~Kononova, A.~Trewartha, K.~Persson,
  G.~Ceder, and A.~Jain, ``Named entity recognition and normalization applied
  to large-scale information extraction from the materials science
  literature,'' \emph{J. Chem. Inf. Model.}, vol.~59, pp. 3692--3702, 2019.

\bibitem{nlp_matsci_mining_0}
\BIBentryALTinterwordspacing
T.~He, W.~Sun, H.~Huo, O.~Kononova, Z.~Rong, V.~Tshitoyan, T.~Botari, and
  G.~Ceder, ``Similarity of precursors in solid-state synthesis as text-mined
  from scientific literature,'' \emph{Chemistry of Materials}, vol.~32, no.~18,
  pp. 7861--7873, 2020. [Online]. Available:
  \url{https://doi.org/10.1021/acs.chemmater.0c02553}
\BIBentrySTDinterwordspacing

\bibitem{nlp_matsci_0}
\BIBentryALTinterwordspacing
K.~Hatakeyama-Sato and K.~Oyaizu, ``Integrating multiple materials science
  projects in a single neural network,'' \emph{Communications Materials},
  vol.~1, no.~1, p.~49, Jul 2020. [Online]. Available:
  \url{https://doi.org/10.1038/s43246-020-00052-8}
\BIBentrySTDinterwordspacing

\bibitem{nlp_matsci_1}
\BIBentryALTinterwordspacing
O.~Kononova, T.~He, H.~Huo, A.~Trewartha, E.~A. Olivetti, and G.~Ceder,
  ``Opportunities and challenges of text mining in materials research,''
  \emph{iScience}, vol.~24, no.~3, p. 102155, 2021. [Online]. Available:
  \url{https://www.sciencedirect.com/science/article/pii/S2589004221001231}
\BIBentrySTDinterwordspacing

\bibitem{nlp_matsci_2}
\BIBentryALTinterwordspacing
E.~A. Olivetti, J.~M. Cole, E.~Kim, O.~Kononova, G.~Ceder, T.~Y.-J. Han, and
  A.~M. Hiszpanski, ``Data-driven materials research enabled by natural
  language processing and information extraction,'' \emph{Applied Physics
  Reviews}, vol.~7, no.~4, p. 041317, 2020. [Online]. Available:
  \url{https://doi.org/10.1063/5.0021106}
\BIBentrySTDinterwordspacing

\bibitem{nlp_matsci_3}
T.~Dieb, M.~Yoshioka, S.~Hara, and M.~Newton, ``Framework for automatic
  information extraction from research papers on nanocrystal devices,''
  \emph{Beilstein J. Nanotechnol.}, vol.~6, pp. 1872--1882, 2015.

\bibitem{nlp_matsci_4}
M.~Gaultois, T.~Sparks, C.~Borg, R.~Seshadri, W.~Bonificio, and D.~Clarke,
  ``Data-driven review of thermoelectric materials: Performance and resource
  considerations,'' \emph{Chem. Mater.}, vol.~25, pp. 2911--2920, 2013.

\bibitem{nlp_chem_transfer}
N.~Pang, L.~Qian, W.~Lyu, and J.-D. Yang, ``Transfer learning for scientific
  data chain extraction in small chemical corpus with bert-crf model,'' 2019.

\bibitem{nlp_chem_ner_0}
P.~Corbett and A.~Copestake, ``Cascaded classifiers for confidence-based
  chemical named entity recognition,'' \emph{BMC Bioinformatics}, vol.~9, no.
  Suppl 11, p.~S4, 2008.

\bibitem{nlp_chem_ner_1}
M.~Krallinger, O.~Rabal, A.~Lourenço, J.~Oyarzabal, and A.~Valencia,
  ``Information retrieval and text mining technologies for chemistry,''
  \emph{Chem. Rev.}, vol. 117, pp. 7673--7761, 2017.

\bibitem{nlp_chem_ner_2}
T.~Rockt{\"{a}}schel, M.~Weidlich, and U.~Leser, ``Chemspot: A hybrid system
  for chemical named entity recognition,'' \emph{Bioinformatics}, vol.~28, pp.
  1633--1640, 2012.

\bibitem{nlp_chem_ner_3}
M.~Krallinger, O.~Rabal, F.~Leitner, M.~Vazquez, D.~Salgado \emph{et~al.},
  ``The chemdner corpus of chemicals and drugs and its annotation principles,''
  \emph{J. Cheminform.}, vol.~7, p.~S2, 2015.

\bibitem{nlp_chem_ner_4}
R.~Leaman, C.-H. Wei, and Z.~Lu, ``tmchem: a high performance approach for
  chemical named entity recognition and normalization,'' \emph{J. Cheminform.},
  vol.~7, p.~S3, 2015.

\bibitem{nlp_chem_ner_5}
I.~Korvigo, M.~Holmatov, A.~Zaikovskii, and M.~Skoblov, ``Putting hands to
  rest: efficient deep cnn-rnn architecture for chemical named entity
  recognition with no hand-crafted rules,'' \emph{J. Cheminform.}, vol.~10,
  p.~28, 2018.

\bibitem{nlp_biomed_ner_0}
M.~Garc{\'\i}a-Remesal, A.~Garc{\'\i}a-Ruiz, D.~P{\'e}rez-Rey, D.~{De La
  Iglesia}, and V.~Maojo, ``Using nanoinformatics methods for automatically
  identifying relevant nanotoxicology entities from the literature,''
  \emph{Biomed. Res. Int.}, vol. 2013, 2013.

\bibitem{matbert_ner}
\BIBentryALTinterwordspacing
A.~Trewartha, N.~Walker, H.~Huo, S.~Lee, K.~Cruse, J.~Dagdelen, A.~Dunn, K.~A.
  Persson, G.~Ceder, and A.~Jain, ``Quantifying the advantage of
  domain-specific pre-training on named entity recognition tasks in materials
  science,'' \emph{Patterns}, vol.~3, no.~4, p. 100488, 2022. [Online].
  Available:
  \url{https://www.sciencedirect.com/science/article/pii/S2666389922000733}
\BIBentrySTDinterwordspacing

\bibitem{tm_mat_disc_0}
\BIBentryALTinterwordspacing
F.~Ren, L.~Ward, T.~Williams, K.~J. Laws, C.~Wolverton, J.~Hattrick-Simpers,
  and A.~Mehta, ``Accelerated discovery of metallic glasses through iteration
  of machine learning and high-throughput experiments,'' \emph{Science
  Advances}, vol.~4, no.~4, p. eaaq1566, 2018. [Online]. Available:
  \url{https://www.science.org/doi/abs/10.1126/sciadv.aaq1566}
\BIBentrySTDinterwordspacing

\bibitem{tm_mat_disc_1}
\BIBentryALTinterwordspacing
C.~C. Fischer, K.~J. Tibbetts, D.~Morgan, and G.~Ceder, ``Predicting crystal
  structure by merging data mining with quantum mechanics,'' \emph{Nature
  Materials}, vol.~5, no.~8, pp. 641--646, Aug 2006. [Online]. Available:
  \url{https://doi.org/10.1038/nmat1691}
\BIBentrySTDinterwordspacing

\bibitem{tm_protocol}
\BIBentryALTinterwordspacing
L.~Weston, V.~Tshitoyan, J.~Dagdelen, O.~Kononova, A.~Trewartha, K.~A. Persson,
  G.~Ceder, and A.~Jain, ``Named entity recognition and normalization applied
  to large-scale information extraction from the materials science
  literature,'' \emph{Journal of Chemical Information and Modeling}, vol.~59,
  no.~9, pp. 3692--3702, Sep 2019. [Online]. Available:
  \url{https://doi.org/10.1021/acs.jcim.9b00470}
\BIBentrySTDinterwordspacing

\bibitem{tm_char_0}
\BIBentryALTinterwordspacing
X.~Wang, J.~Li, H.~D. Ha, J.~C. Dahl, J.~C. Ondry, I.~Moreno-Hernandez,
  T.~Head-Gordon, and A.~P. Alivisatos, ``Autodetect-mnp: An unsupervised
  machine learning algorithm for automated analysis of transmission electron
  microscope images of metal nanoparticles,'' \emph{JACS Au}, vol.~1, no.~3,
  pp. 316--327, Mar 2021. [Online]. Available:
  \url{https://doi.org/10.1021/jacsau.0c00030}
\BIBentrySTDinterwordspacing

\bibitem{tm_char_1}
\BIBentryALTinterwordspacing
N.~J. Szymanski, C.~J. Bartel, Y.~Zeng, Q.~Tu, and G.~Ceder, ``Probabilistic
  deep learning approach to automate the interpretation of multi-phase
  diffraction spectra,'' \emph{Chemistry of Materials}, vol.~33, no.~11, pp.
  4204--4215, Jun 2021. [Online]. Available:
  \url{https://doi.org/10.1021/acs.chemmater.1c01071}
\BIBentrySTDinterwordspacing

\bibitem{nano_db}
\BIBentryALTinterwordspacing
X.~Yan, A.~Sedykh, W.~Wang, B.~Yan, and H.~Zhu, ``Construction of a web-based
  nanomaterial database by big data curation and modeling friendly
  nanostructure annotations,'' \emph{Nature Communications}, vol.~11, no.~1, p.
  2519, May 2020. [Online]. Available:
  \url{https://doi.org/10.1038/s41467-020-16413-3}
\BIBentrySTDinterwordspacing

\bibitem{aunp_synth_dataset}
\BIBentryALTinterwordspacing
K.~Cruse, A.~Trewartha, S.~Lee, Z.~Wang, H.~Huo, T.~He, O.~Kononova, A.~Jain,
  and G.~Ceder, ``Text-mined dataset of gold nanoparticle synthesis procedures,
  morphologies, and size entities,'' \emph{Scientific Data}, vol.~9, no.~1, p.
  234, May 2022. [Online]. Available:
  \url{https://doi.org/10.1038/s41597-022-01321-6}
\BIBentrySTDinterwordspacing

\bibitem{s2s}
\BIBentryALTinterwordspacing
I.~Sutskever, O.~Vinyals, and Q.~V. Le, ``Sequence to sequence learning with
  neural networks,'' 2014. [Online]. Available:
  \url{https://arxiv.org/abs/1409.3215}
\BIBentrySTDinterwordspacing

\bibitem{gpt3}
T.~Brown, B.~Mann, N.~Ryder, M.~Subbiah, J.~D. Kaplan, P.~Dhariwal,
  A.~Neelakantan, P.~Shyam, G.~Sastry, A.~Askell \emph{et~al.}, ``Language
  models are few-shot learners,'' \emph{Advances in neural information
  processing systems}, vol.~33, pp. 1877--1901, 2020.

\bibitem{gpt3_mastci}
\BIBentryALTinterwordspacing
A.~Dunn, J.~Dagdelen, N.~Walker, S.~Lee, A.~S. Rosen, G.~Ceder, K.~Persson, and
  A.~Jain, ``Structured information extraction from complex scientific text
  with fine-tuned large language models,'' 2022. [Online]. Available:
  \url{https://arxiv.org/abs/2212.05238}
\BIBentrySTDinterwordspacing

\bibitem{sol_synth_dataset}
\BIBentryALTinterwordspacing
Z.~Wang, O.~Kononova, K.~Cruse, T.~He, H.~Huo, Y.~Fei, Y.~Zeng, Y.~Sun, Z.~Cai,
  W.~Sun, and G.~Ceder, ``Dataset of solution-based inorganic materials
  synthesis recipes extracted from the scientific literature,'' 2021. [Online].
  Available: \url{https://doi.org/10.48550/arXiv.2111.10874}
\BIBentrySTDinterwordspacing

\bibitem{aunpdataset}
\BIBentryALTinterwordspacing
K.~Cruse, A.~Trewartha, S.~Lee, Z.~Wang, H.~Huo, T.~He, O.~Kononova, A.~Jain,
  and G.~Ceder, ``Text-mined aunp synthesis recipes dataset,'' figshare, 2021.
  [Online]. Available: \url{https://doi.org/10.6084/m9.figshare.16614262.v3}
\BIBentrySTDinterwordspacing

\bibitem{gpt1}
A.~Radford and K.~Narasimhan, ``Improving language understanding by generative
  pre-training,'' 2018.

\bibitem{gpt2}
A.~Radford, J.~Wu, R.~Child, D.~Luan, D.~Amodei, and I.~Sutskever, ``Language
  models are unsupervised multitask learners,'' 2019.

\bibitem{example_text}
\BIBentryALTinterwordspacing
M.~Ma, H.~Chen, Y.~Chen, X.~Wang, F.~Chen, X.~Cui, and J.~Shi, ``Au capped
  magnetic core/mesoporous silica shell nanoparticles for combined
  photothermo-/chemo-therapy and multimodal imaging,'' \emph{Biomaterials},
  vol.~33, no.~3, pp. 989--998, 2012. [Online]. Available:
  \url{https://www.sciencedirect.com/science/article/pii/S0142961211012208}
\BIBentrySTDinterwordspacing

\bibitem{example}
\BIBentryALTinterwordspacing
K.~W. Smith, H.~Zhao, H.~Zhang, A.~S{\'a}nchez-Iglesias, M.~Grzelczak, Y.~Wang,
  W.-S. Chang, P.~Nordlander, L.~M. Liz-Marz{\'a}n, and S.~Link, ``Chiral and
  achiral nanodumbbell dimers: The effect of geometry on plasmonic
  properties,'' \emph{ACS Nano}, vol.~10, no.~6, pp. 6180--6188, Jun 2016.
  [Online]. Available: \url{https://doi.org/10.1021/acsnano.6b02194}
\BIBentrySTDinterwordspacing

\bibitem{synth_kbh4}
\BIBentryALTinterwordspacing
M.~Zareie, X.~Xu, and M.~Cortie, ``In situ organization of gold nanorods on
  mixed self-assembled-monolayer substrates,'' \emph{Small}, vol.~3, no.~1, pp.
  139--145, 2007. [Online]. Available:
  \url{https://onlinelibrary.wiley.com/doi/abs/10.1002/smll.200600280}
\BIBentrySTDinterwordspacing

\bibitem{ar_lspr_0}
X.~Huang, S.~Neretina, and M.~A. El-Sayed, ``\BIBforeignlanguage{en}{Gold
  nanorods: from synthesis and properties to biological and biomedical
  applications},'' \emph{\BIBforeignlanguage{en}{Adv Mater}}, vol.~21, no.~48,
  pp. 4880--4910, Jul. 2009.

\bibitem{ar_lspr_1}
\BIBentryALTinterwordspacing
L.~Vigderman and E.~R. Zubarev, ``High-yield synthesis of gold nanorods with
  longitudinal spr peak greater than 1200 nm using hydroquinone as a reducing
  agent,'' \emph{Chemistry of Materials}, vol.~25, no.~8, pp. 1450--1457, Apr
  2013. [Online]. Available: \url{https://doi.org/10.1021/cm303661d}
\BIBentrySTDinterwordspacing

\bibitem{ar_range}
\BIBentryALTinterwordspacing
L.~Feng, Z.~Xuan, J.~Ma, J.~Chen, D.~Cui, C.~Su, J.~Guo, and Y.~Zhang,
  ``Preparation of gold nanorods with different aspect ratio and the optical
  response to solution refractive index,'' \emph{Journal of Experimental
  Nanoscience}, vol.~10, no.~4, pp. 258--267, 2015. [Online]. Available:
  \url{https://doi.org/10.1080/17458080.2013.824619}
\BIBentrySTDinterwordspacing

\bibitem{growth_limits}
\BIBentryALTinterwordspacing
N.~D. Burrows, S.~Harvey, F.~A. Idesis, and C.~J. Murphy, ``Understanding the
  seed-mediated growth of gold nanorods through a fractional factorial design
  of experiments,'' \emph{Langmuir}, vol.~33, no.~8, pp. 1891--1907, Feb 2017.
  [Online]. Available: \url{https://doi.org/10.1021/acs.langmuir.6b03606}
\BIBentrySTDinterwordspacing

\bibitem{limited_growth}
``Hcl-retarded gold nanorod growth for aspect ratio and shape tuning,''
  \emph{Journal of Nanoscience and Nanotechnology}, vol.~16, no.~1, 2016.

\bibitem{hcl_growth_0}
Y.~Wang, Y.~Guo, Y.~Shen, R.~Chen, F.~Wang, D.~Zhou, and S.~Guo,
  ``\BIBforeignlanguage{en}{{HCl-Retarded} gold nanorod growth for aspect ratio
  and shape tuning},'' \emph{\BIBforeignlanguage{en}{J Nanosci Nanotechnol}},
  vol.~16, no.~1, pp. 1194--1201, Jan. 2016.

\bibitem{hcl_growth_1}
\BIBentryALTinterwordspacing
M.-Z. Wei, T.-S. Deng, Q.~Zhang, Z.~Cheng, and S.~Li, ``Seed-mediated synthesis
  of gold nanorods at low concentrations of ctab,'' \emph{ACS Omega}, vol.~6,
  no.~13, pp. 9188--9195, Apr 2021. [Online]. Available:
  \url{https://doi.org/10.1021/acsomega.1c00510}
\BIBentrySTDinterwordspacing

\bibitem{gpt3_data}
\BIBentryALTinterwordspacing
``{GPT}-3 seed-mediated {A}u{NR} synthesis extraction datasets,'' 2023.
  [Online]. Available: \url{https://doi.org/10.6084/m9.figshare.19719310.v2}
\BIBentrySTDinterwordspacing

\bibitem{xsede}
\BIBentryALTinterwordspacing
J.~Towns, T.~Cockerill, M.~Dahan, I.~Foster, K.~Gaither, A.~Grimshaw,
  V.~Hazlewood, S.~Lathrop, D.~Lifka, G.~D. Peterson, R.~Roskies, J.~R. Scott,
  and N.~Wilkins-Diehr, ``Xsede: Accelerating scientific discovery,''
  \emph{Computing in Science \& Engineering}, vol.~16, no.~5, pp. 62--74,
  Sept.-Oct. 2014. [Online]. Available:
  \url{doi.ieeecomputersociety.org/10.1109/MCSE.2014.80}
\BIBentrySTDinterwordspacing

\end{thebibliography}
\end{document}